\documentclass[journal]{IEEEtai}

\usepackage[colorlinks,urlcolor=blue,linkcolor=blue,citecolor=blue]{hyperref}

\usepackage{color,array}

\usepackage{amsmath}
\usepackage{amssymb}
\usepackage{graphicx}   
\usepackage{subcaption}
\usepackage{wrapfig}
\usepackage{float}

\usepackage{booktabs}
\usepackage{multirow}
\usepackage{adjustbox}
\usepackage{colortbl}
\usepackage{xcolor}

\usepackage{hyperref}
\usepackage{cite}
\usepackage{comment}

\definecolor{bestcell}{HTML}{E8F5E9}
\usepackage{rotating}

\begin{document}

\title{T-Backdoor: Exploiting Temporal Redundancy in Neuromorphic Data for Spike-preserving Backdoor Attacks on SNNs}

\author{
    \IEEEauthorblockN{Abdullah Arafat Miah\IEEEauthorrefmark{1}, Kevin Vu\IEEEauthorrefmark{1}, Yu Bi\IEEEauthorrefmark{1}}\\
    \IEEEauthorblockA{\IEEEauthorrefmark{1}Department of Electrical, Computer, and Biomedical Engineering\\
    University of Rhode Island, Kingston, RI, USA\\
    \{abdullaharafat.miah, kevin\_vu029, yu\_bi\}@uri.edu}
}

\maketitle

\begin{abstract}

Backdoor attacks are a serious security threat to deep neural networks (DNNs) and remain largely underexplored for spiking neural networks (SNNs). Existing attacks primarily introduce spatiotemporal triggers that induce deviations in the spike distribution of poisoned samples relative to their clean counterparts. To address this limitation, this work proposes a novel backdoor attack on SNNs, termed \textbf{T-Backdoor}, which operates using purely temporal triggers such as \textit{Rate}, \textit{Latency}, and \textit{Jitter} without introducing any spatial perturbation, making the shift in spike distributions significantly harder to detect. Through extensive experiments on three benchmark neuromorphic datasets: N-MNIST, CIFAR10-DVS, and N-Caltech101, and evaluation against seven baseline backdoor defense methods, we demonstrate that T-Backdoor achieves a near-perfect 100\% attack success rate (ASR) in both single target and multi target settings with only minor degradation in clean accuracy, while remaining robust against existing backdoor detection and mitigation techniques. The codes are available at https://github.com/SiSL-URI/T-Backdoor .

\end{abstract}



\section{Introduction}

\label{sec:intro}


Spiking neural networks (SNNs) have been regarded as energy-efficient alternatives to the conventional deep neural networks (DNNs) \cite{eshraghian2023training}. DNNs work with continuous activations, while SNNs work through discrete spike events, making them consume less energy on neuromorphic hardware \cite{davies2018loihi}. These properties make SNNs suitable for processing neuromorphic data which are processed by dynamic vision sensors (DVS)\cite{serrano2013dvs}. While binary activation is a challenge for backpropagation due to its non-differentiability, recent advances in surrogate gradient training \cite{neftci2019surrogate} fill the gaps to achieve accuracy comparable to DNNs. SNNs have been adopted in various safety-critical applications such as autonomous driving \cite{chen2020event} \cite{viale2021carsnn} and medical diagnosis \cite{kasabov2014evolving}.

Despite the advances, the security implications of deploying SNNs remain largely underexplored. Backdoor attacks have emerged as one of the serious security threats to DNNs in which an adversary embeds a hidden trigger utilizing data poisoning, and the model wrongly classifies the triggered inputs to an attacker defined labels. While backdoor attacks are extensively studied in deep neural networks \cite{gu2017badnets} \cite{nguyen2021wanet} \cite{miah2024noiseattack}, research on backdoor attacks on SNNs with neuromorphic data is still in its infancy. Existing works such as sneaky spikes \cite{abad2023sneaky}, propose spatiotemporal triggers that inject spatial patches into event streams, which optionally modulate their location a cross time frames. While effective, these triggers shift the spike distribution of the poisoned data from the clean counterparts and can be detected with simple spike count statistics such as Kullback--Leibler (KL) divergence, Wasserstein distance, and total spike residual. Such shortcoming motivates the central question of this work: \emph{can an adversary compromise an SNN without perturbing the spike distribution of the poisoned samples?} To answer this question, we propose \textbf{T-Backdoor}, a temporal-only backdoor attacks against SNNs trained on neuromorphic data. The temporal dimension of the neuromorphic data constitutes an additional attack surface which is exploited in this work by proposing three triggers: (i) \emph{Rate Trigger}, which modulates the temporal playback speed of the event stream by a scaling factor, (ii) \emph{Latency Trigger}, which introduces a fixed delay in the temporal axis, and (iii) \emph{Jitter Trigger}, which swaps a configurable number of frame pairs to induce controlled temporal reordering. As these triggers work purely within the temporal axis without adding any spatial content, they preserve the spike distribution of the poisoned samples. Our contributions can be summarized as follows:


\begin{itemize}
    \item We identify and investigate a fundamental drawback of existing spatiotemporal backdoor triggers on SNNs: their addition to the spatial contents shifts the spike distribution of the poisoned samples, rendering them detectable by simple spike-count statistics. 
    \item We propose a novel backdoor attack on SNNs named \textbf{T-Backdoor}, the first purely temporal backdoor attack framework for SNNs encompassing three spike-preserving trigger types: Rate, Latency, and Jitter. For the first time of backdoor attacks on SNNs, we further investigate and showcase that T-Backdoor is scalable effectively to multi-target payload scenarios.
    \item Through extensive experiments on N-MNIST, CIFAR10-DVS and N-Caltech101, we show that T-Backdoor achieves near-perfect attack success rates (ASR $\approx 100\%$) with negligible clean accuracy degradation, evades spike-distribution-based detection, and resists seven state-of-the-art backdoor detection and mitigation techniques.
\end{itemize}

\vspace{-2mm}


\section{Related Works}
\label{sec:related}

Backdoor attacks have been studied in various deep learning domains, such as convolutional neural networks and vision transformers \cite{gu2019badnets, nguyen2021wanet}, large language models \cite{chen2021badnl, miah2024exploiting, li2021hidden}, and graph neural networks \cite{xi2021graph, khan2026multi, zeng2021rethinking}. BadNet \cite{gu2017badnets} was the first attack to introduce backdoor attacks in image classification tasks, where an adversary poisons the training dataset to modify the parameters of a DNN so that it maps a patch-based trigger to an attacker-defined target label. Subsequent research has focused on innovative trigger design, such as the blend attack \cite{chen2017targeted}, in which a universal trigger is blended with the image to activate backdoors. To increase the stealthiness of the trigger, reflection-based triggers \cite{liu2020reflection}, learnable imperceptible noise-based triggers \cite{doan2021lira}, wrapping-based triggers \cite{nguyen2021wanet}, and quantization-based triggers \cite{wang2022bppattack} have been proposed. Furthermore, frequency-based triggers \cite{feng2022fiba, gao2024dual, zeng2021rethinking} have been proposed to hide triggers in the frequency domain. Besides input-space triggers, feature-based backdoor attacks have also been proposed \cite{doan2021backdoor, cheng2021deep}. While in most backdoor attacks the training ground truth is altered to embed the backdoor behavior, clean-label attacks preserve the training ground truth and insert the backdoor into the model by only modifying the target-label samples \cite{turner2018clean, zhao2020clean, zeng2023narcissus}. For backdoor attacks on language models, the triggers can be characters, words, sentences, homographs, linguistic styles, and more \cite{chen2021badnl, li2021hidden, pan2022hidden, yang2021careful}.

To defend against backdoor attacks, the defense categories can be divided into two main types: backdoor detection and backdoor mitigation. In backdoor detection, several characteristics of a backdoor attack can be detected, such as backdoor trigger detection, backdoored model detection, and target label detection using trigger reverse-engineering optimization, as in Neural Cleanse \cite{wang2019nc} and ABS \cite{liu2019abs}, as well as poisoned sample detection, as in STRIP \cite{gao2019strip} and TeCo \cite{liu2023detecting}. For mitigating backdoor attacks, several methods have been proposed: fine-tuning-based approaches \cite{lin2024unveiling, zhu2023enhancing}, where the backdoored model is fine-tuned with limited training data to remove the backdoor effect from the model; backdoor neuron pruning-based approaches \cite{wu2021adversarial, liu2018finepruning, li2023reconstructive, miah2026deepcurer}, where the neurons in a backdoored model that significantly influence the backdoor behavior are pruned to mitigate the backdoor effect; and black-box approaches \cite{yang2024sampdetox, miah2026lite, shi2023black}, where poisoned samples are purified so that the trigger does not exist in the input to activate the backdoor behavior.

Backdoor research on spiking network is comparatively scarce. Sneaky spikes \cite{abad2024sneaky} is the first neuromorphic backdoor attack on SNNs proposing multiple spatiotemporal triggers. Ria{\~n}o et al.~\cite{riano2024flashy} extends this line of work to physical-world attacks using LED-based triggers captured by real DVS cameras. In BadSNN \cite{miah2026badsnn} the hyperparameters of the spiking neurons are varied during backdoor training to induce backdoors in the model. The backdoor attack is also proposed in spiking federated learning \cite{abad2025time} \cite{fu2024spikewhisper}. All of these attacks are based on spatiotemporal triggers. On the defense side, \cite{li2025unsupervised} proposes an unsupervised backdoor detection and mitigation framework for neuromorphic backdoors in SNNs, which utilizes the maximum margin statistic of the final spiking layer's temporal membrane potential to detect the backdoored model and target label, and a mitigation technique that clamps the weights connecting the first two convolutional layers.


\section{Preliminaries}

SNNs compute discrete spikes using spiking neurons, most commonly based on the leaky integrate-and-fire (LIF) neuron model \cite{maass1997networks} expressed by the governing equation \ref{eq:lif}. Here $V[t]$ represents the membrane potential of the neuron at time step $t$, $\alpha \in (0, 1]$ serves as a temporal decay factor. The terms $S_i[t]$ and $w_i$ correspond to the $i^{th}$ input spike and its respective synaptic weight. The firing threshold is denoted by $\vartheta$, while $O[t]$ represents the neuron's output, determined by the Heaviside step function $u(\cdot)$. Whenever $V[t]$ surpasses the threshold $\vartheta$, the neuron emits a spike, setting $O[t] = 1$.


\begin{equation}
V[t] = \alpha V[t-1] + \sum_{i} w_i S_i[t] - \vartheta O[t-1],\;
O[t] = u(V[t] - \vartheta)
\label{eq:lif}
\end{equation}


\noindent SNNs commonly operate on neuromorphic data, a time-encoded representation of illumination changes captured by a dynamic vision sensor (DVS) camera. Each event can be formally described as a tuple $e = (x, y, t, p)$, where $(x, y)$ represents the spatial coordinate of the pixel, $t$ denotes the precise timestamp of the event, and $p \in \{+1, -1\}$ indicates the polarity of the brightness change. A positive polarity, referred to as an ON event, corresponds to an increase in brightness, whereas a negative polarity, known as an OFF event, corresponds to a decrease. These two polarities are typically visualized using distinct colors to differentiate the direction of intensity change.

\section{Spike Distribution Analysis of the Spatiotemporal Triggers} \label{subsec:spike_distribution_analysis_sneaky_spikes}
\begin{figure*}
    \centering
    \includegraphics[width=\linewidth]{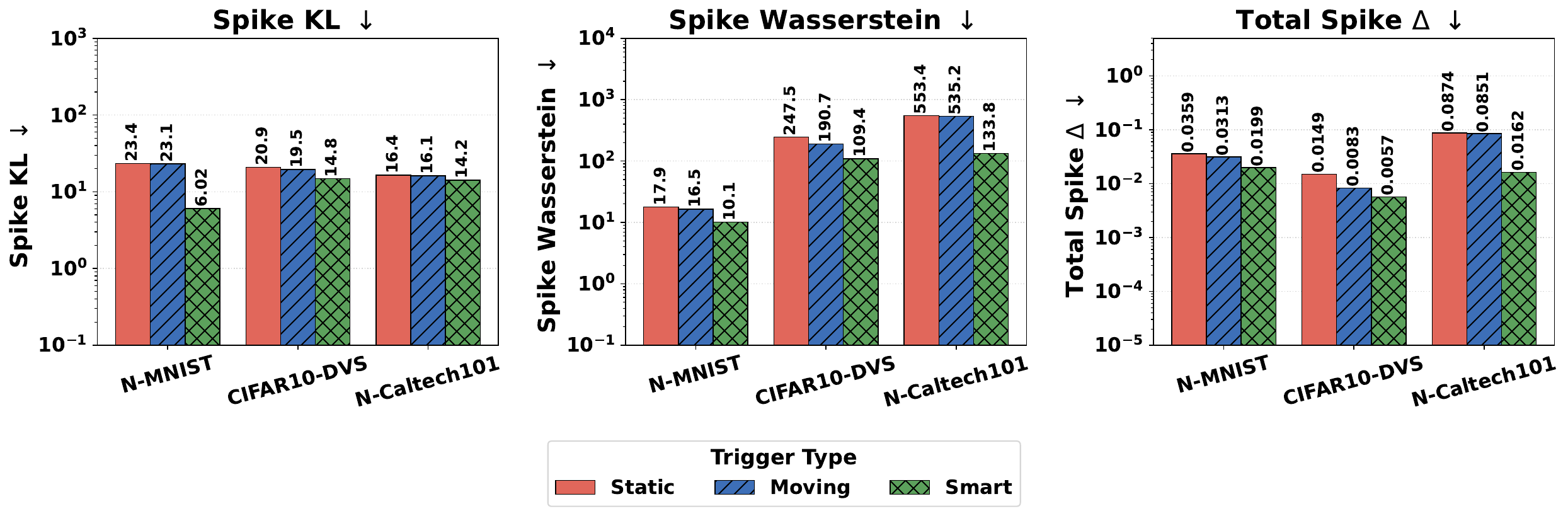}
    \caption{Spike distribution distances and residual between clean and poisoned samples of the spatiotemporal triggers.}
    \label{fig:preli1}
\end{figure*}


\noindent In this section, we investigate the divergence between the spike distribution of clean and poisoned samples for three spatiotemporal triggers: Static, Moving, and Smart proposed in \cite{abad2023sneaky}. We take 1000 random samples from three widely used neuromorphic datasets: N-MNIST \cite{orchard2015converting}, CIFAR10-DVS \cite{li2017cifar10dvs}, and N-Caltech101 \cite{orchard2015converting}, and measure three divergence metrics: spike Kullback-Leibler (KL) divergence, spike Wasserstein distance, and total spike delta ($\Delta$). Spike KL divergence is a measure of how one probability of spike distribution differs from a reference distribution. Let $\mathbf{p} = (p_1, \dots, p_T)$ and $\mathbf{q} = (q_1, \dots, q_T)$ denote the normalized frame-wise spike count distributions of the clean and poisoned samples, respectively. The KL divergence can be expressed as $D_{\mathrm{KL}}(\mathbf{p} \| \mathbf{q}) = \sum_{t=1}^{T} p_t \log \frac{p_t}{q_t + \epsilon},$ where $\epsilon$ is a small constant for numerical stability. We also compute the Wasserstein-1 (earth mover's) distance between the frame-wise spike count sequences of clean and poisoned samples which can be expressed $W_1(\mathbf{p}, \mathbf{q}) = \inf_{\gamma \in \Gamma(\mathbf{p}, \mathbf{q})} \sum_{t, t'} |t - t'| \, \gamma(t, t'),$ where $\Gamma(\mathbf{p}, \mathbf{q})$ is the set of all couplings between $\mathbf{p}$ and $\mathbf{q}$. Finally, we compute changes in the total spike count between the clean and poisoned samples using $\Delta = \frac{|S_{\mathrm{poison}} - S_{\mathrm{clean}}|}{S_{\mathrm{clean}} + \epsilon},$ where $S = \sum_{t=1}^{T} \sum_{c,h,w} x_{t,c,h,w}$ is the total spike count. The measure of these distance metrics are given in the Figure \ref{fig:preli1}.  On N-MNIST, Static achieves a spike KL of 23.45 and a spike Wasserstein distance of 17.94, while Moving and Smart produce similarly elevated values. On CIFAR10-DVS, the Wasserstein distances are substantially higher, 247.50 for Static and 190.67 for Moving, reflecting the pronounced spike distribution anomalies that spatiotemporal triggers introduce on higher-resolution event streams. N-Caltech101 follows a similar trend, with Static and Moving reaching Wasserstein distances above 500. The Smart trigger occupies an intermediate position: its adaptive placement strategy reduces some of the spike distribution perturbation relative to Static and Moving, yet it still introduces detectable spike count changes (e.g., spike KL of 6.02 on N-MNIST versus 23.45 for Static). \textit{These scores highlight a drawback in the previous spatiotemporal-based triggers that poisoned samples intrinsically exhibit a different spike distribution in comparison with the clean counterparts.} 

\section{Proposed Attack} \label{sec:methodology}
\subsection{Threat Model}
In the attack threat model, we assume that the attacker has access to the training dataset and the model, which is consistent with prior work \cite{abad2023sneaky, gu2017badnets}. Since we propose a data poisoning technique to insert a backdoor into the victim model, the adversary can also be viewed as a dataset provider without direct access to the model. The attacker can be a third-party provider who trains the SNN on behalf of the victim, who outsources it due to limited resources. Another real-world setting is that the attacker trains the model and releases the backdoored model in third-party repositories, where the victim uses the model unknowingly. Specifically, we utilize data poisoning and a dirty-label methodology to inject backdoors into SNNs. We investigate the proposed backdoor attack in the digital image domain.

\subsection{Temporal Triggers Design}

\begin{figure}[!htbp]
    \centering
    \includegraphics[width=1.05\linewidth]{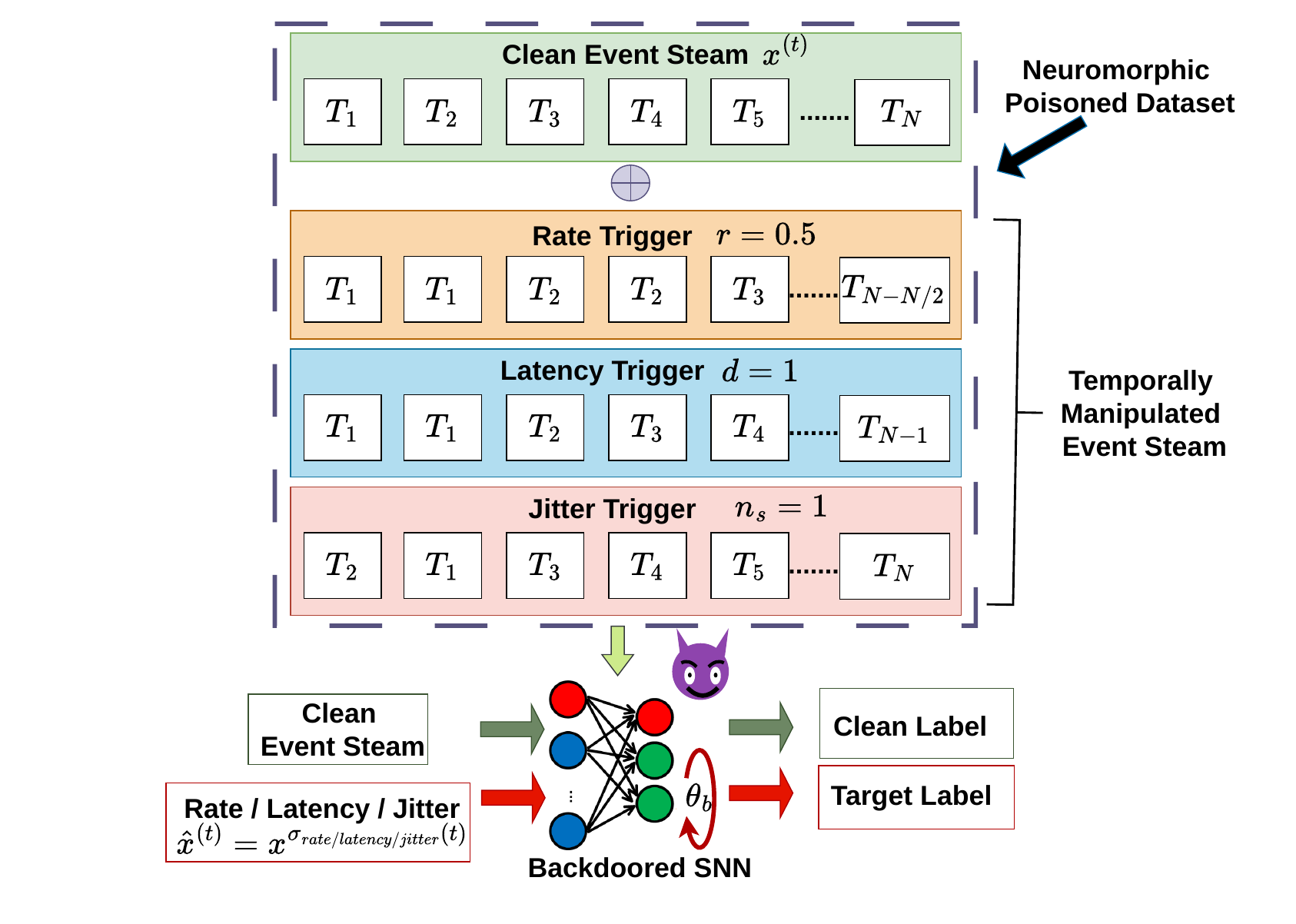}
    \caption{Overview of the proposed T-Backdoor in which we propose three purely temporal triggers: Rate, Latency, and Jitter that preserve the spatial contents of the poisoned samples, keeping the spike distribution close to the clean counterparts.}
    \label{fig:overview}
\end{figure}

\begin{figure}[h]
    \centering

    \begin{subfigure}{\linewidth}
        \centering
        \includegraphics[width=\linewidth]{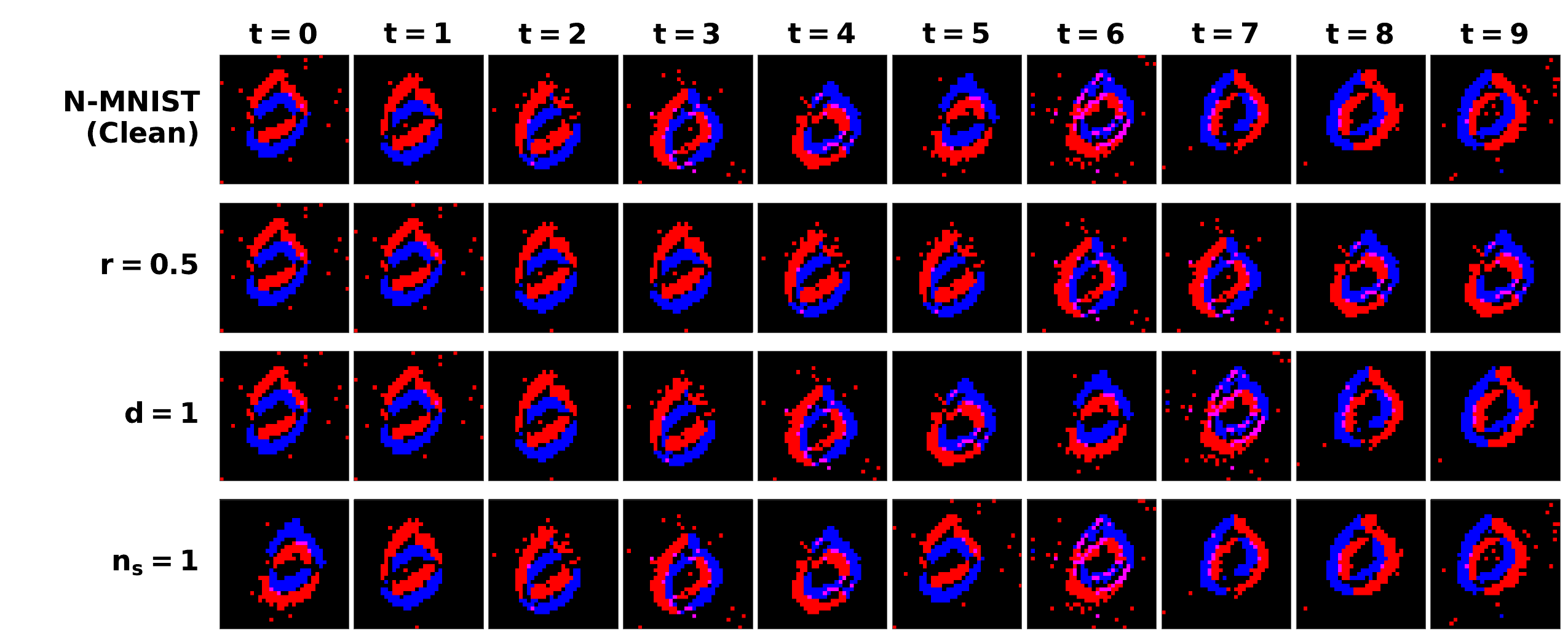}
        \caption{N-MNIST}
    \end{subfigure}

    \vspace{2mm}

    \begin{subfigure}{\linewidth}
        \centering
        \includegraphics[width=\linewidth]{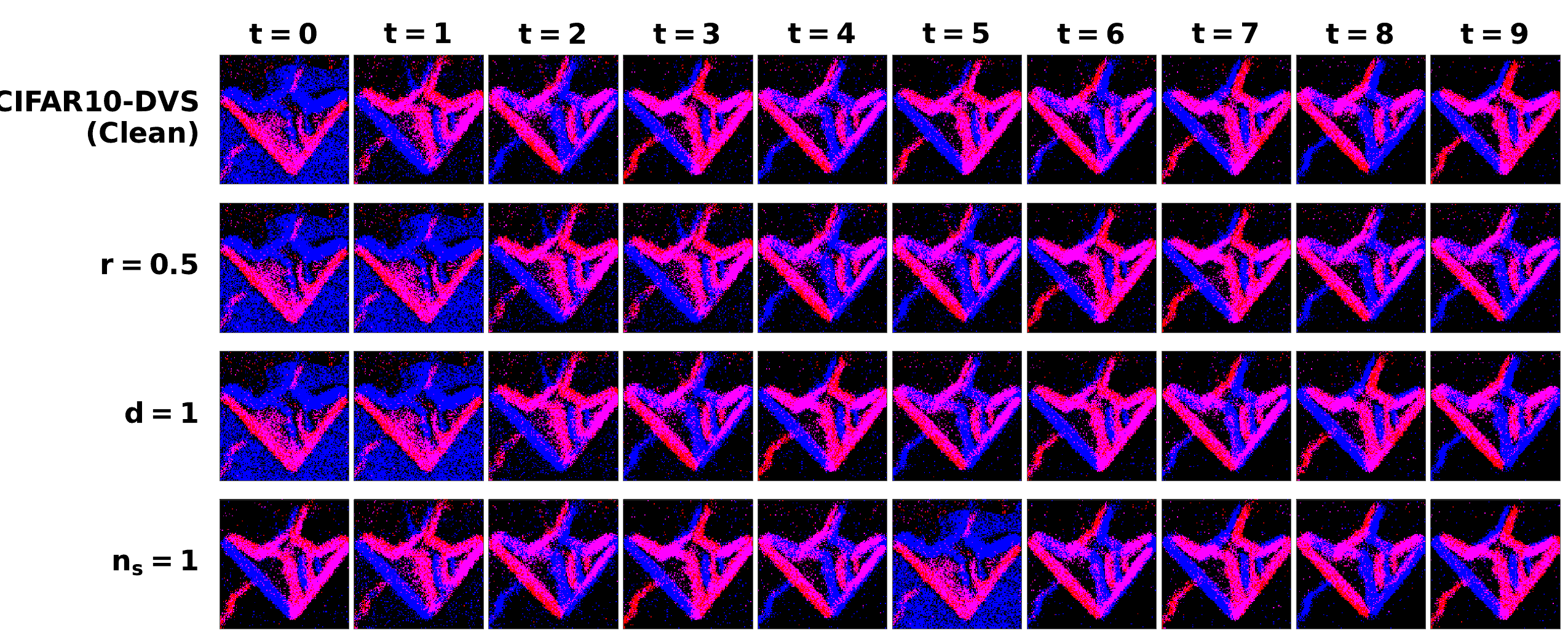}
        \caption{CIFAR10-DVS}
    \end{subfigure}

    \vspace{2mm}

    \begin{subfigure}{\linewidth}
        \centering
        \includegraphics[width=\linewidth]{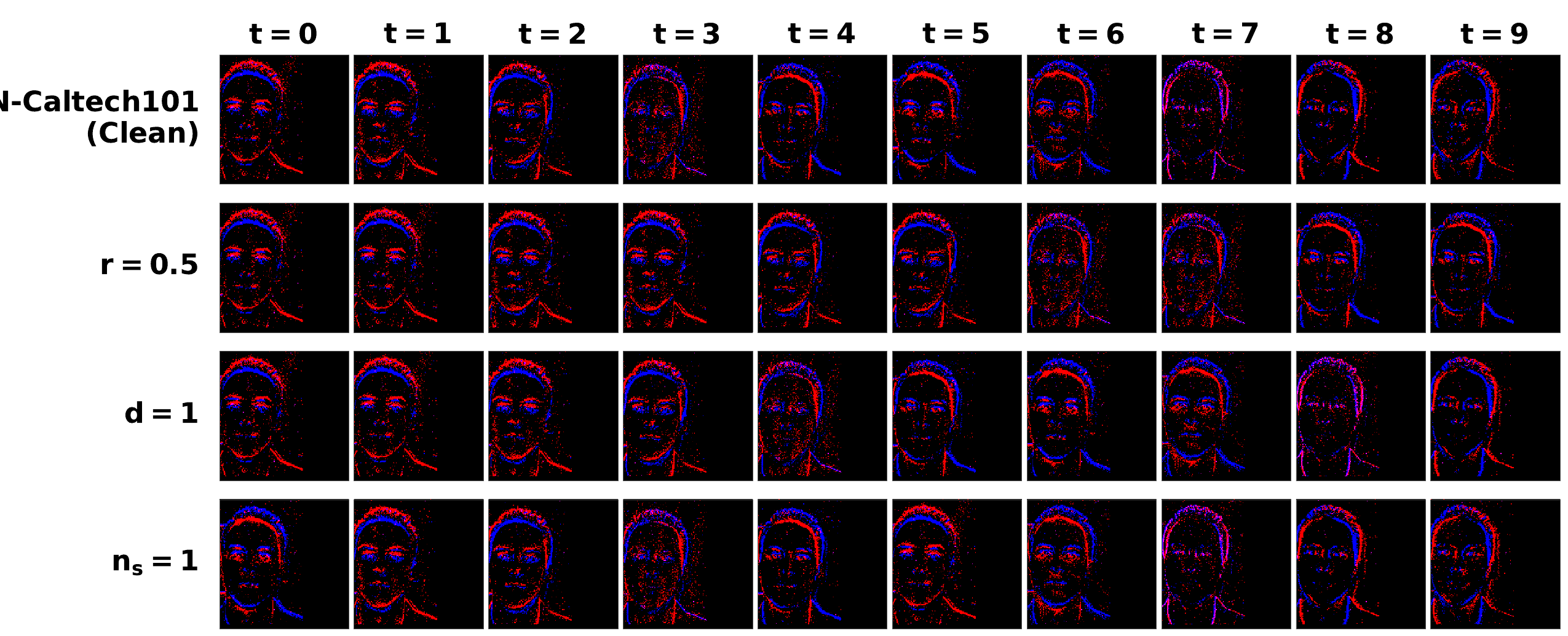}
        \caption{N-Caltech101}
    \end{subfigure}

    \caption{Visualization of temporal triggers on all three benchmark datasets}
    \label{fig:all_triggers}
\end{figure}

Neuromorphic data are typically encoded into $T$ time frames and $p$ polarities. The temporal dimension $T$ introduces an additional attack surface for adversaries, enabling the design of purely temporal triggers, which differ from previously studied spatiotemporal trigger designs~\cite{abad2023sneaky}. As illustrated in Figure \ref{fig:overview}, we propose three types of temporal triggers: \textbf{Rate}, \textbf{Latency}, and \textbf{Jitter}, each exploiting a different aspect of the temporal dynamics of neuromorphic data. We formalize all three temporal triggers under a unified index-remapping framework.
Let $\mathbf{x} = \{\mathbf{x}^{(0)}, \mathbf{x}^{(1)}, \ldots, \mathbf{x}^{(T-1)}\}$ denote a clean neuromorphic sample with $T$ frames. A temporal trigger is defined as a deterministic remapping function $\sigma : \{0, \ldots, T{-}1\} \to \{0, \ldots, T{-}1\}$ that produces a triggered sample $\hat{\mathbf{x}}$ by reassigning frame indices: $\hat{\mathbf{x}}^{(t)} = \mathbf{x}^{(\sigma(t))}, \quad t = 0, 1, \ldots, T{-}1.$ The visualization of each triggers is available in Figure~\ref{fig:all_triggers}.

 \noindent \textbf{Rate Trigger.} The Rate trigger essentially modulates the speed (aka. temporal playback rate) of the neuromorphic data. Given a scaling factor $r > 0$, the rate trigger function rescales the frame indices to adjust the temporal playback rate, which is then learned by the victim model as a backdoor pattern. Such manipulation alters only the temporal dynamics of the neuromorphic data, without modifying any spatial structure or spike distribution. For a clean sample $\mathbf{x}$ and a time step $t$ where $t = {0, \ldots, T{-}1}$, Rate trigger function can be mathematically expressed as $\hat{\mathbf{x}}^{(t)} = \mathbf{x}^{(\sigma_{\text{rate}}(t))}, \quad \sigma_{\text{rate}}(t) = \min\!\bigl(\lfloor r \cdot t \rfloor,\; T{-}1\bigr),$ where $\lfloor \cdot \rfloor$ denotes the floor function, and $\min$ clamps the rescaled index to the valid range ${0, \ldots, T{-}1}$. Based on the scaling factor $r$, Rate trigger exhibits two distinct characteristics: speed-up ($r > 1$) and slow-down ($r < 1$). When $r > 1$, $\sigma_{\text{rate}}$ advances through the original events faster than real time, effectively skipping intermediate events. The speed-up operation can be expressed by Equation~\ref{eq:rate_speedup}.

\vspace{-1mm}

\begin{equation}
    \hat{\mathbf{x}}^{(t)} = 
    \begin{cases}
        \mathbf{x}^{(\lfloor r \cdot t \rfloor)}, & \text{if } \lfloor r \cdot t \rfloor < T, \\[4pt]
        \mathbf{x}^{(T-1)},                        & \text{otherwise}.
    \end{cases}
    \label{eq:rate_speedup}
\end{equation}


\noindent For $r < 1$, $\sigma_{\text{rate}}$ advances through the original events more slowly than real time by repeating individual frames across multiple consecutive output time steps. The slow-down operation can be expressed by Equation~\ref{eq:rate_slowdown}.


\begin{equation}
    \hat{\mathbf{x}}^{(t)} = \mathbf{x}^{(\lfloor r \cdot t \rfloor)}, \quad \text{where } \lfloor r \cdot t \rfloor \leq \lfloor r \cdot (T{-}1) \rfloor < T{-}1.
    \label{eq:rate_slowdown}
\end{equation}


\noindent \textbf{Latency Trigger.} The Latency trigger introduces a temporal delay into the dynamics of the neuromorphic data by shifting the entire event sequence forward by $d$ time steps, where $d$ is the delay factor. In this trigger scenario, the sensor or a preprocessing pipeline induces latency in the temporal dynamics, which is then learned by the victim model as a backdoor pattern. For a given $\mathbf{x}$ and a time step $t$ where $t = {0, \ldots, T{-}1}$, Latency triggered sample can be expressed by Equation~\ref{eq:latency}. The delay factor $d$ controls the strength and stealthiness of the trigger. A smaller $d$ (e.g. $d$ = 1 or 2) produces a subtle shift where a larger $d$ may cause significant temporal displacement. The triggered sample is then $\hat{\mathbf{x}}^{(t)} = \mathbf{x}^{(\sigma_{\text{lat}}(t))}$ for all $t$.

\vspace{-3mm}
 
\begin{equation}
     \hat{\mathbf{x}}^{(t)} = 
    \begin{cases}
        \mathbf{x}^{(0)}, & \text{if } t < d, \\[4pt]
        \mathbf{x}^{(t - d)}, & \text{if } d \leq t < T,
    \end{cases}
    \label{eq:latency}
\end{equation}


\noindent \textbf{Jitter Trigger.} The Jitter trigger controlled temporal displacement by swapping specific pairs of frames, which is fixed and remains consistent across all poisoned samples. By reordering the temporal structure, we do not require any addition or deletion of event sequences making the overall spike statistics identical to the clean counterparts. Let $\mathcal{P} = \{(i_1, j_1), (i_2, j_2), \ldots, (i_{n_s}, j_{n_s})\}$ denote a set of ${n_s}$ swap pairs, where each $(i_{n_s}, j_{n_s})$ specifies two distinct frame indices to be exchanged, with $i_{n_s} \neq j_{n_s}$ and $i_{n_s}, j_{n_s} \in \{0, \ldots, T{-}1\}$. For a stealthy perturbation, $n_s = 1$ swaps a single pair of frames. Given $n_s$ desired swaps, we compute a step size $s = \max\!\bigl(1, \lfloor T / (2n_s) \rfloor\bigr)$ and construct pairs as: $(i_k, j_k) = (2ks,\; 2ks + s), \quad k = 0, 1, \ldots, n_s{-}1, \quad \text{subject to } j_k < T.$ The resulting remapping $\sigma_{\text{jit}} : \{0, \ldots, T{-}1\} \to \{0, \ldots, T{-}1\}$ is initialized as the identity permutation and modified by applying each swap. The triggered sample is then $\hat{\mathbf{x}}^{(t)} = \mathbf{x}^{(\sigma_{\text{jit}}(t))}$ for all $t$.
\begin{equation}
    \sigma_{\text{jit}}(t) =
    \begin{cases}
        j_k, & \text{if } t = i_k \text{ for some } k, \\
        i_k, & \text{if } t = j_k \text{ for some } k, \\
        t,   & \text{otherwise}.
    \end{cases}
    \label{eq:jitter_mapping}
\end{equation}

\subsection{Backdoor Training} \label{subsec:backdoor_training}

Following the existing works \cite{gu2017badnets, abad2023sneaky}, we design the backdoor training mechanism using the dirty-label data poisoning technique. Given a training dataset $D_{train}$, we can split the training data based on the attack setting. In this work we consider three attack settings: the conventional single-target attack setting (one-to-one), the single-/multi-trigger multi-target attack setting (one-to-N), and the sample-specific attack setting. For every attack setting, the training dataset $D_{train}$ can be split into two parts: $D_{train}^* = D_{train}^{clean} \cup D_{train}^{poison}$. For input sample space $\mathcal{X}$, clean label space $\mathcal{Y}_c$, and target label space $\mathcal{Y}_t$, we can define
\begin{equation}
D_{train}^{clean} = \{(x_i, y_i) : x_i \in \mathcal{X},\; y_i \in \mathcal{Y}_c\},
\label{eq:clean_split}
\end{equation}
\begin{equation}
D_{train}^{poison} = \{(\hat{\mathbf{x}}^{(t)}_i, y_i) : \hat{\mathbf{x}}^{(t)}_i \in \mathcal{X},\; y_i \in \mathcal{Y}_t\},
\label{eq:poison_split}
\end{equation}
where $i = 1,\ldots,N$ denotes the total number of samples.

For the multi-target setting, a distinct target label is assigned to each trigger or different parameters of a specific trigger (i.e. different $r, d, j$ for each rate, latency, and jitter trigger respectively), enabling a one-to-N mapping between triggers and target classes. Let $\mathcal{T} = \{1,\ldots,T\}$ denote the set of trigger indices, where $T$ is the total number of triggers used. We define a target label assignment function $\phi: \mathcal{T} \rightarrow \mathcal{Y}_t$ that maps each trigger index $l \in \mathcal{T}$ to its corresponding target label $y_t \in \mathcal{Y}_t$. The poisoned dataset for the multi-target setting can then be defined as
\begin{equation}
D_{train}^{poison} = \bigcup_{t \in \mathcal{T}} \{(\hat{\mathbf{x}}^{(t)}_i, y_t) : \hat{\mathbf{x}}^{(t)}_i \in \mathcal{X},\; y_t = \phi(l) \in \mathcal{Y}_t\}.
\label{eq:multitarget_split}
\end{equation}

For the sample-specific setting, the training dataset consists of three parts: clean samples with clean labels, triggered samples from victim classes with dirty (target) labels, and triggered samples from non-victim classes that retain their clean labels. Let $\mathcal{Y}_v \subset \mathcal{Y}_c$ denote the set of victim classes. The clean subset is defined as
\begin{equation}
D_{train}^{clean} = \{(x_i, y_i) : x_i \in \mathcal{X},\; y_i \in \mathcal{Y}_c\},
\label{eq:samplespecific_clean}
\end{equation}
the poisoned subset from victim classes as
\begin{equation}
D_{train}^{poison,v} = \{(\hat{\mathbf{x}}_i, y_t) : \hat{\mathbf{x}}_i \in \mathcal{X},\; y_i \in \mathcal{Y}_v,\; y_t \in \mathcal{Y}_t\},
\label{eq:samplespecific_poison_victim}
\end{equation}
and the triggered subset from non-victim classes, which retains clean labels, as
\begin{equation}
D_{train}^{poison,nv} = \{(\hat{\mathbf{x}}_i, y_i) : \hat{\mathbf{x}}_i \in \mathcal{X},\; y_i \in \mathcal{Y}_c \setminus \mathcal{Y}_v\}.
\label{eq:samplespecific_poison_nonvictim}
\end{equation}
The overall training dataset is then given by $D_{train}^* = D_{train}^{clean} \cup D_{train}^{poison,v} \cup D_{train}^{poison,nv}$. Training or fine-tuning an SNN with $D_{train}^{*}$ will induce a backdoor effect in the model corresponding to the desired attack setting. The poison ratio can be defined as $\frac{|D_{train}^{poison}|}{|D_{train}^{*}|}$. Note that for the multi-target setting, the poison ratio is defined with respect to the entire $D_{train}^{poison}$, not with respect to the poisoned samples of each individual target.


\section{Experiments} \label{sec:experiment}
\subsection{Experimental Settings}
\label{sec:exp_settings}
\textbf{Datasets.} To demonstrate the effectiveness of \textbf{T-Backdoor}, we evaluate on three widely used neuromorphic benchmark datasets: N-MNIST~\cite{orchard2015converting}, CIFAR10-DVS~\cite{li2017cifar10dvs}, and N-Caltech101~\cite{orchard2015converting}. We mainly use number of timesteps $T = 10$ for each datasets for our experiments.

\noindent \textbf{Network Architectures.} In terms of network architecture for N-MNIST, we use \texttt{NMNISTNet} \cite{abad2024sneaky} comprises a single convolutional layer and a fully connected layer. For CIFAR10-DVS, we use \texttt{CIFAR10DVSNet} \cite{abad2024sneaky}, which has two convolutional layers, two fully connected layers and voting layers. For N-Caltech101 we use SEW ResNet-18 \cite{fang2021deep}. All models are trained directly with surrogate gradients using the SpikingJelly framework~\cite{fang2023spikingjelly} in multi-step mode with LIF neurons. All models are trained using the Adam optimizer~\cite{kingma2014adam}. Clean baseline accuracies are 99.32\% (N-MNIST), 71.00\% (CIFAR10-DVS), and 78.37\% (N-Caltech101).

\noindent \textbf{Baselines.} We compare the proposed T-Backdoor against three spatiotemporal backdoor attacks proposed in sneaky spikes \cite{abad2024sneaky}: Static, Moving, and Smart. We adopt original hyperparameters reported in the paper: trigger size $10\%$ of the input dimension, polarity $= 1$ for Static and Moving, and $c=2$ masks for Smart. 

\noindent \textbf{Defense Evaluation.} To demonstrate the robustness of T-Backdoor, we evaluate seven representative backdoor defenses. We utilize temporal membrane potential backdoor detection (TMPBD)~\cite{li2025tmpbd}, the most recent SNN specific backdoor detection method. To evaluate the backdoor mitigation, we utilize three fine-tuning-based methods: vanilla fine tuning, neural attention distillation (NAD) \cite{li2021nad} and two-stage backdoor defense (TSBD)~\cite{lin2024unveiling} and three backdoor neuron pruning based methods: adversarial neuron pruning (ANP)~\cite{wu2021adversarial} with pruning thresholds $thr \in \{0.2, 0.5\}$, channel Lipschitz pruning (CLP)~\cite{zheng2022data}, and fine-pruning~\cite{liu2018finepruning}. We use the official code repository released by the corresponding authors, and adapt exclusively the necessary components to support the SNN architectures employed in our experiments. All mitigation defenses are given access to a clean set of 500 samples for N-MNIST and 100 samples for both CIFAR10-DVS and N-Caltech101.

\noindent \textbf{Evaluation Metrics.} We report two standard metrics for backdoor attacks: i) Clean Accuracy (CA): the accuracy of the backdoored model on clean samples, ii) Attack Success Rate (ASR): the accuracy of the backdoored model on poisoned samples. Class '0' is used for all datasets as a default target label for our experiments. 

\subsection{Attack Effectiveness Analysis}

\begin{figure}[!htbp]
\centering
\includegraphics[width=\linewidth]{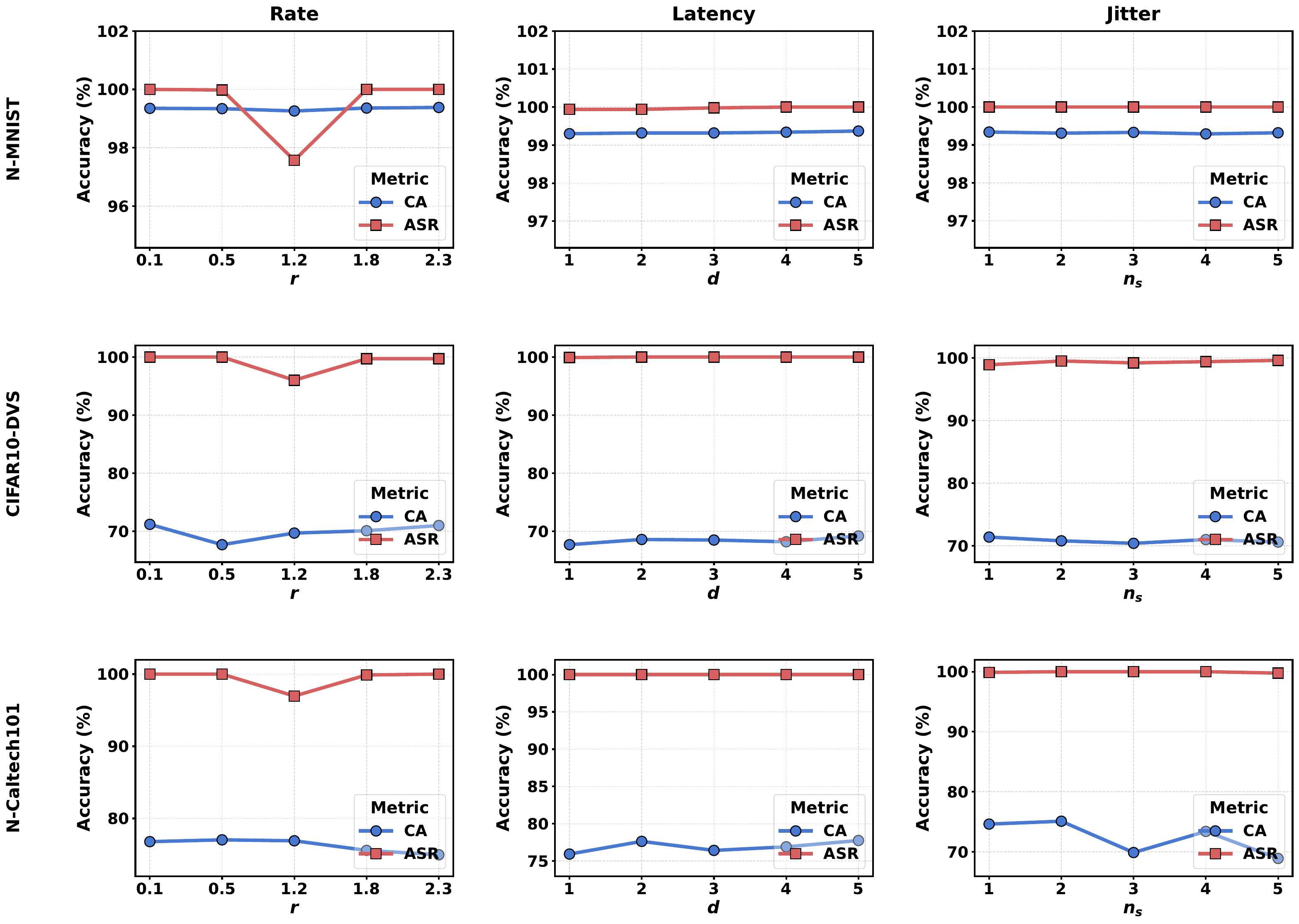}
\caption{Single-target backdoor attack results across datasets with Rate, Latency, and Jitter triggers at $p = 0.1$. Each trigger varies only its own parameter: Rate varies scale factor $r$; Latency varies delay $d$; Jitter varies swap pairs $n_s$. Clean-model CA: 99.32\% (N-MNIST), 71.00\% (CIFAR10-DVS), 78.37\% (N-Caltech101).}
\label{fig:single_target}
\end{figure}

In this experiment, we analyze the effectiveness of the proposed T-Backdoor with its three trigger types: Rate, Latency, and Jitter across all three neuromorphic datasets: N-MNIST, CIFAR10-DVS and N-Caltech101. For each dataset, the trigger parameters of each trigger type are varied independently. The scaling factor $r$ is varied across 0.1, 0.5, 1.2, 1.8, and 2.3 for Rate, while the delay frames $d$ and number of frame pairs to swap $n_s$ are each varied from 1 to 5 for Latency and Jitter. As presented in Figure~\ref{fig:single_target}, ASRs of the Rate trigger reach $\approx$100\% for every dataset and scaling factor of $r$ with the exception of $r = 1.2$, where it drops to 97.57\% for N-MNIST, 96.00\% for CIFAR10-DVS and 96.96\% for N-Caltech101. Since $r = 1.2$ is the value closest to the neutral scaling factor of 1, the modified spike firing rate closely resembles the original, which may introduce ambiguous temporal features that reduce the model's confidence in associating the trigger pattern with the target class, resulting in a marginal ASR drop. No meaningful drop of model's clean accuracy is observed for N-MNIST. As for CIFAR10-DVS, CA decreases by approximately 1\% in most cases, with slightly larger drops of $\sim$3\% at $r = 0.5$ and $\sim$2\% at $r = 1.2$, both of which remain negligible. Similarly, the CA drop of N-Caltech101 evaluation falls within the negligible 1\%--3\% range across all $r$ values. Comparably, near-perfect ASR of $\approx$100\% is achieved across all datasets and all values of $d$ for the Latency trigger. CA remains unaffected for N-MNIST, while it drops by an average of $\sim$3\% for CIFAR10-DVS and by $\sim$2\% for N-Caltech101, all of which are within an acceptable range. As for the Jitter trigger experiments, near-perfect ASR of $\approx$100\% is also achieved across all datasets and $n_s$ variations. No CA drop is observed for N-MNIST, while only a negligible drop is seen for CIFAR10-DVS. However, the impact on CA for N-Caltech101 is more nuanced: the drop at $n_s \in \{1, 2\}$ is negligible in approximation of 2\%--3\%, yet CA at $n_s \geq 3$ degrades more substantially with drops of $\sim$7\% at $n_s = 3$, $\sim$4\% at $n_s = 4$, and $\sim$9\% at $n_s = 5$. Such observation suggests that aggressive temporal reordering may disrupt the model's ability to learn discriminative features on a high-complexity dataset such as N-Caltech101. Consequently, Jitter achieves better CA preservation and maintains stealthiness at lower values of $n_s$. \textit{In summary, all three temporal triggers: Rate, Latency, and Jitter demonstrate high ASR while preserving CA at levels comparable to an uncompromised baseline, making T-Backdoor both potent and stealthy in terms of model utility degradation. Among the three trigger designs, Latency is the most stable trigger, maintaining near-perfect ASR with minimal CA impact across all configurations. The Rate trigger is sensitive to values of $r$ close to 1, and Jitter performs best and is most stealthy at lower values of $n_s$.}

\vspace{-3mm}

\subsection{Spike Distribution Analysis of T-Backdoor}

\begin{figure*} [!htbp]
    \centering
    \includegraphics[width=\linewidth]{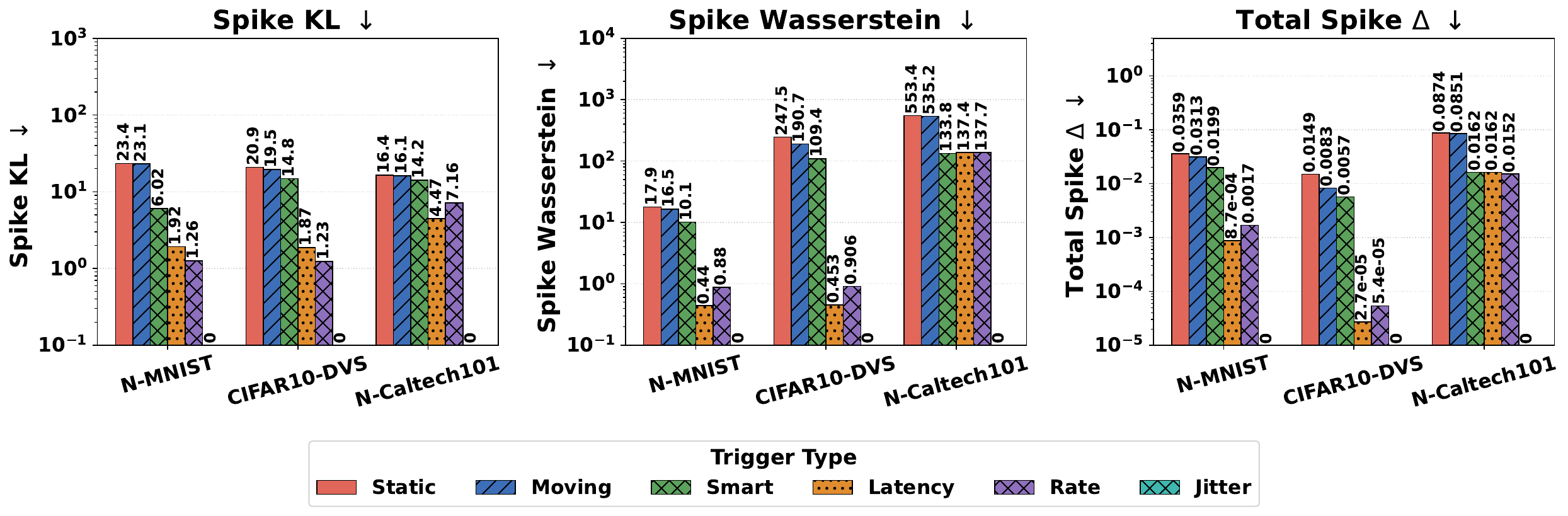}
    \caption{Spike distribution distances and residual between clean and poisoned samples for spatiotemporal and temporal triggers.}
    \label{fig:spike_histogram}
\end{figure*}

\noindent \textbf{Spike Distribution Metrics.} In this section, we analyze the spike distribution of poisoned samples against their clean counterparts across all three neuromorphic datasets: N-MNIST, CIFAR10-DVS and N-Caltech101. The rationale is straightforward: if a poisoned sample exhibits an anomalous spike distribution relative to clean data, such distributional shift can work as a detection mechanism. Indeed, spike count analysis constitutes a natural and model-agnostic detection strategy for neuromorphic data, as it requires no access to trained network parameters. We compare our proposed temporal triggers against the spatiotemporal triggers introduced in sneaky spikes. Presented in Figure~\ref{fig:spike_histogram}, a clear dichotomy emerges between spatiotemporal and temporal triggers. As stated in section \ref{subsec:spike_distribution_analysis_sneaky_spikes} the spatiotemporal triggers exhibit higher values in all three spike distribution metrics: spike KL divergence, spike Wasserstein distance, and total spike residual.
In contrast, the temporal triggers in our proposed T-Backdoor exhibit near-zero perturbation across all three metrics. On N-MNIST, Latency achieves a Spike KL of merely 1.92 with a Total Spike~$\Delta$ of 0.0009, while Rate yields 1.26 and 0.0017, respectively. A similar trend holds for CIFAR10-DVS, where the gap between spatial and temporal triggers is even more pronounced, as well as for N-Caltech101, which shows a moderate but consistent difference. Most notably, Jitter achieves exactly 0.0 on all three metrics across all three datasets. It is due to the fact that Jitter operates entirely through event reordering without introducing any changes in the spatial domain, rendering it intrinsically invisible to spike statistics-based detection methods.

\begin{figure}[!htbp]
    \centering
    \begin{subfigure}[b]{\linewidth}
        \centering
        \includegraphics[width=\linewidth]{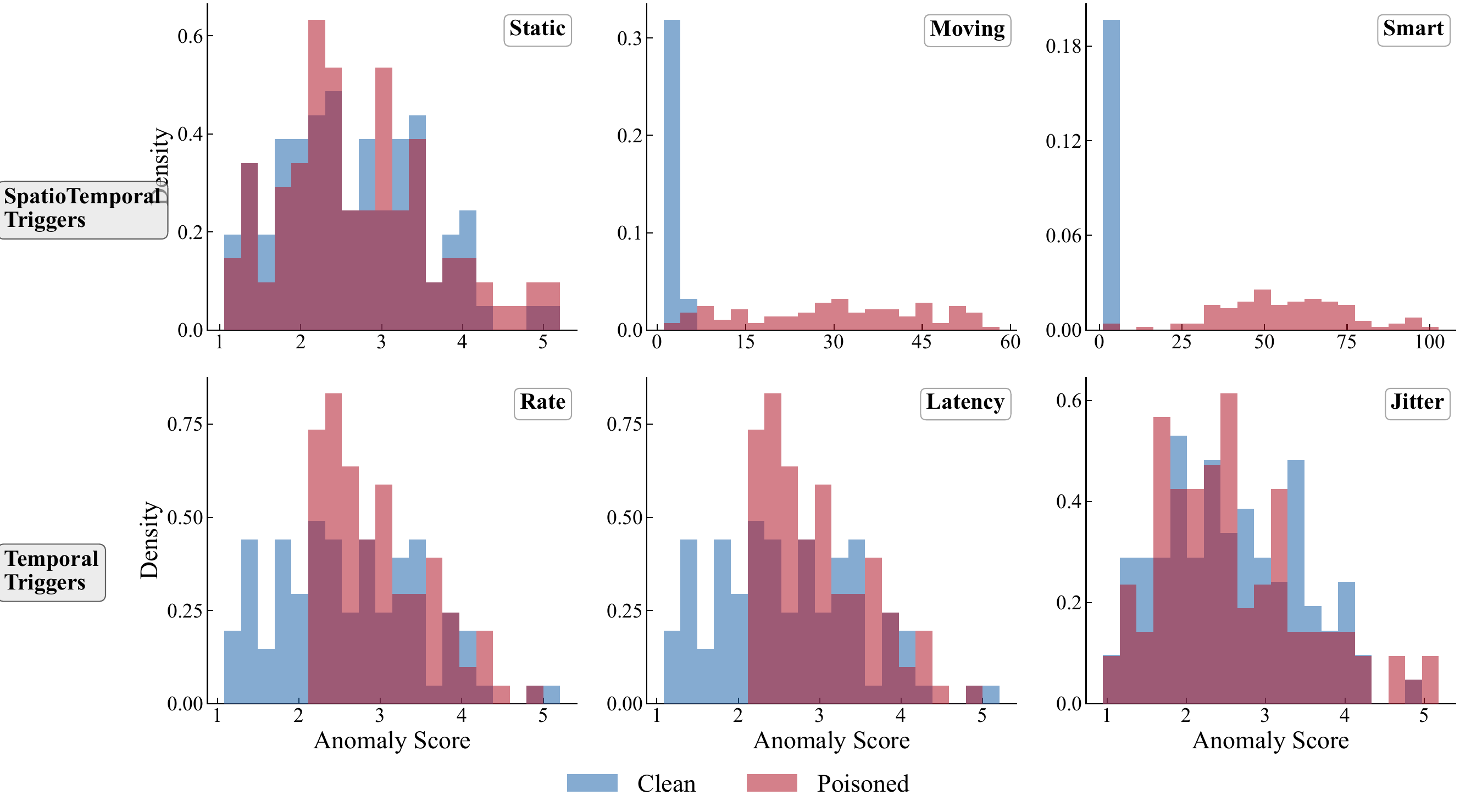}
        \caption{N-MNIST}
        \label{fig:nmnist}
    \end{subfigure}
    
    
    \begin{subfigure}[b]{\linewidth}
        \centering
        \includegraphics[width=\linewidth]{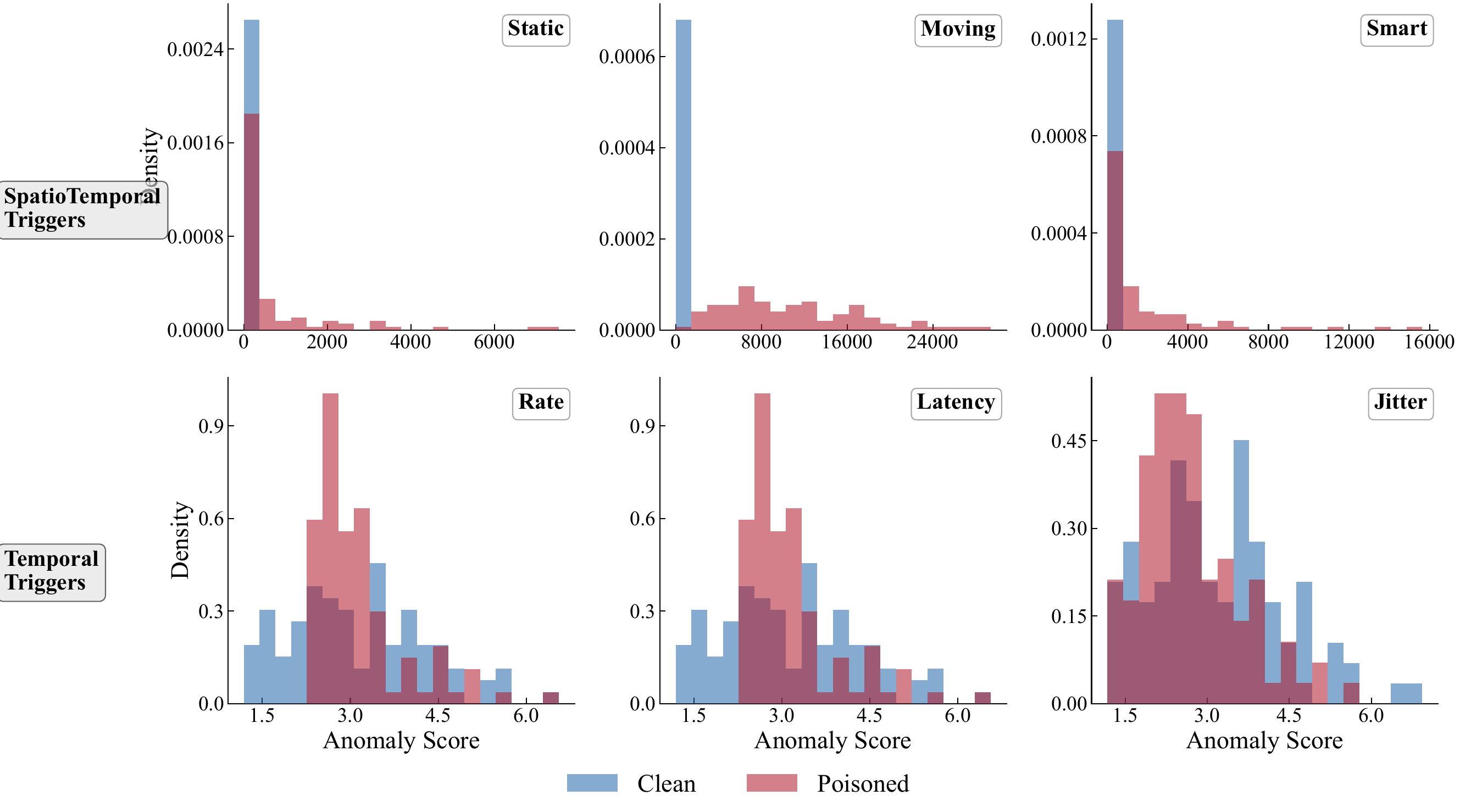}
        \caption{CIFAR-10 DVS}
        \label{fig:cifar10_dvs}
    \end{subfigure}
    
    
    \begin{subfigure}[b]{\linewidth}
        \centering
        \includegraphics[width=\linewidth]{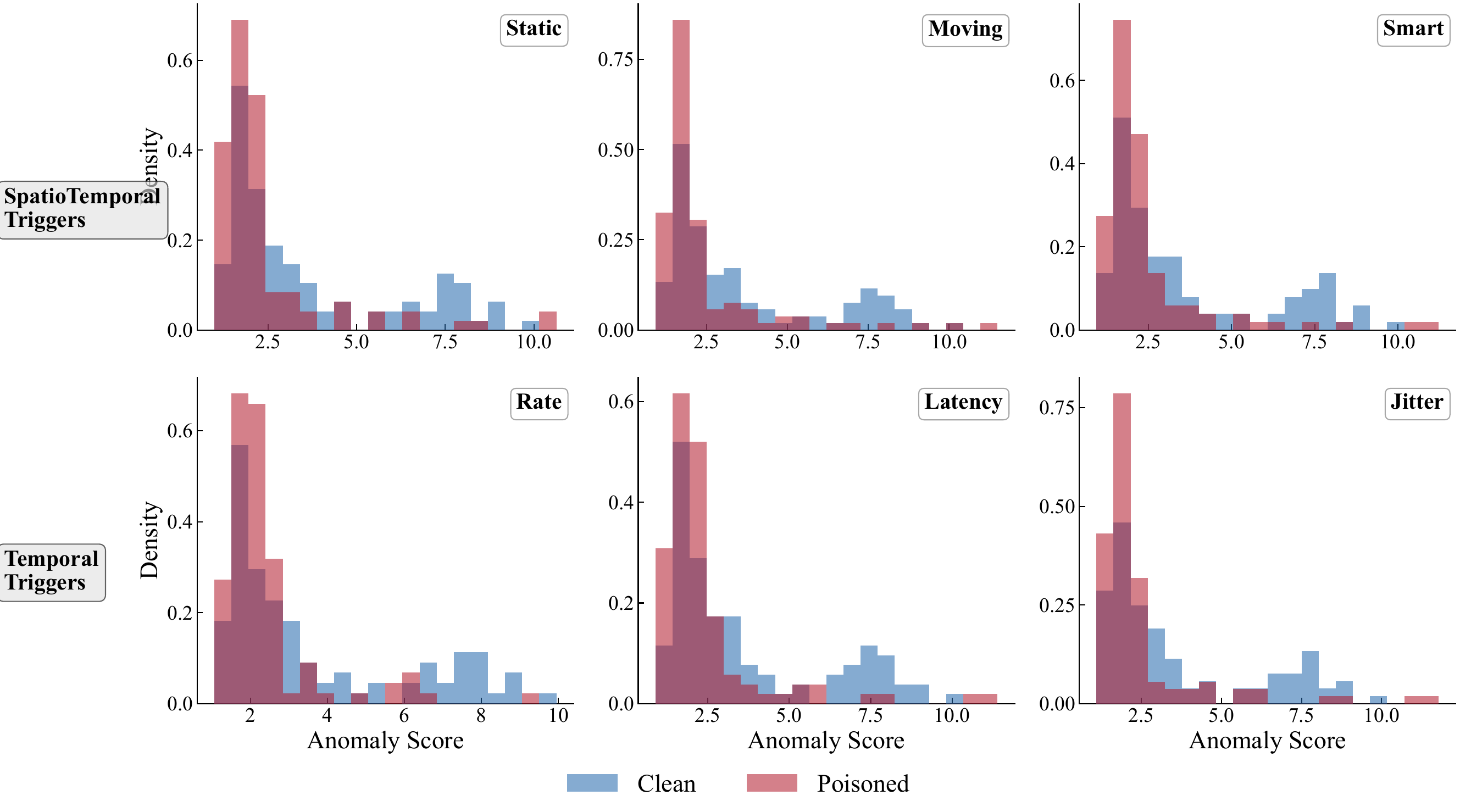}
        \caption{NCALTECH-101}
        \label{fig:ncaltech101}
    \end{subfigure}
    
    \caption{Spike anomaly score distribution of all three datasets.}
    \label{fig:Spike_Anomoly_Score_all}
\end{figure}

\noindent \textbf{Spike Anomaly Score Distributions.} To further evaluate the distributional shift, we conduct a spike anomaly score distribution analysis. To calculate the spike anomaly score, we take a reference set of clean samples and compute four component scores for each test sample: Total Spike Count Score~$s_{\mathrm{total}}$, Frame-wise Spike Distribution Score~$s_{\mathrm{frame}}$, Inter-frame Difference Score~$s_{\mathrm{inter}}$, and Frame Variance Score~$s_{\mathrm{var}}$. Let $\mu_S$ and $\sigma_S$ denote the mean and standard deviation of total spike counts across the clean reference set. As for a test sample with total spike count~$S$, the Total Spike Count Score~$s_{\mathrm{total}}$ is given in Equation~\ref{eq:anomaly_score}. Let $\mathbf{f} = (f_1, \dots, f_T)$ be the per-frame spike counts of the test sample, and let $\mathbf{f}_{\mathrm{ref}}$ be the concatenated frame-wise spike counts across all clean reference samples. The Frame-wise Spike Distribution Score~$s_{\mathrm{frame}}$ is the normalized Wasserstein-1 distance, expressed in Equation~\ref{eq:anomaly_score}, where $\bar{f}_{\mathrm{ref}}$ is the mean of $\mathbf{f}_{\mathrm{ref}}$. We also compute the absolute inter-frame differences $\mathbf{d} = (|f_2 - f_1|, \dots, |f_T - f_{T-1}|)$ and compare them against the reference inter-frame differences~$\mathbf{d}_{\mathrm{ref}}$ using the same normalized Wasserstein distance. Regarding the Frame Variance Score, we compute the variance of per-frame spike counts $v = \mathrm{Var}(\mathbf{f})$ of the test sample, and let $\mu_v$ and $\sigma_v$ be the mean and standard deviation of this quantity across clean references. Finally, the Spike Anomaly Score~$\mathcal{A}$ is obtained as the equally-weighted sum of all four components, as defined in Equation~\ref{eq:anomaly_score}. We use 100 clean samples as the reference set and plot the anomaly score distributions of 1{,}000 clean samples alongside their poisoned counterparts. Anomaly score distributions of all three datasets are presented in Figure~\ref{fig:Spike_Anomoly_Score_all}.

\begin{equation}
\resizebox{\linewidth}{!}{$
\begin{aligned}
s_{\mathrm{total}} &= \frac{|S - \mu_S|}{\sigma_S}, \quad
s_{\mathrm{frame}} = \frac{W_1(\mathbf{f}, \mathbf{f}_{\mathrm{ref}})}{\bar{f}_{\mathrm{ref}} + \epsilon}, \quad
s_{\mathrm{inter}} = \frac{W_1(\mathbf{d}, \mathbf{d}_{\mathrm{ref}})}{\bar{d}_{\mathrm{ref}} + \epsilon}, \\
s_{\mathrm{var}} &= \frac{|v - \mu_v|}{\sigma_v},  \quad
\mathcal{A} = s_{\mathrm{total}} + s_{\mathrm{frame}} + s_{\mathrm{inter}} + s_{\mathrm{var}}.
\end{aligned}
$}
\label{eq:anomaly_score}
\end{equation}

\begin{figure}[!htbp]
\centering
\includegraphics[width=\linewidth]{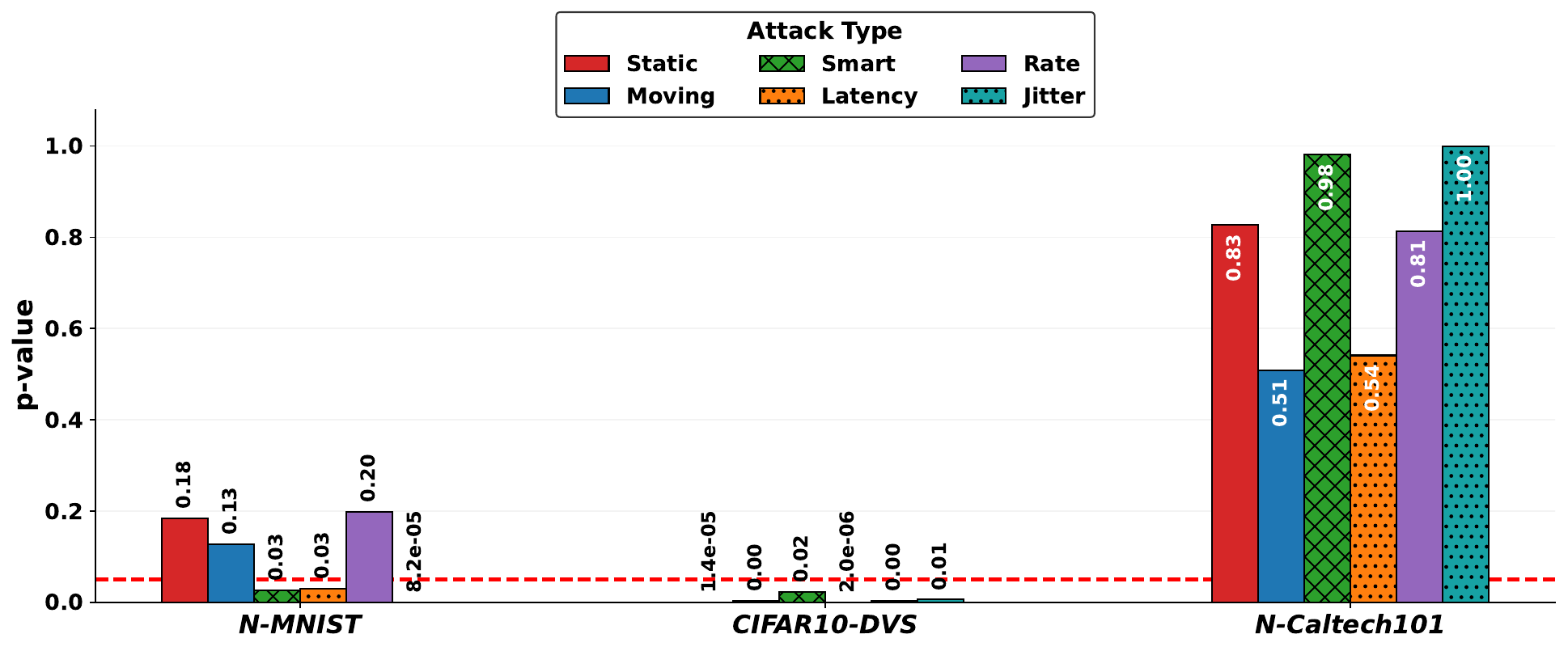}
\caption{Backdoor detection using TMPBD. The red dot marks the $p$-value threshold; models with lower $p$-values are flagged as backdoored.}
\label{fig:tmpbd}
\end{figure}

\noindent It can be seen in Figure~\ref{fig:Spike_Anomoly_Score_all} that the spatiotemporal triggers produce clear separation between clean and poisoned distributions on N-MNIST: the Static trigger shows moderate separation with the poisoned distribution shifted rightward, while Moving and Smart generate dramatic separation with poisoned scores extending far beyond the clean range. Such separation is even more pronounced on CIFAR10-DVS, where clean samples produce scores concentrated near zero while all three spatiotemporal triggers yield anomaly scores extending into the thousands. On the contrary, the temporal triggers produce distributions that are nearly indistinguishable from clean ones, with Jitter's poisoned distribution aligning almost perfectly with the clean baseline. As for N-Caltech101, the spatiotemporal triggers exhibit substantial overlap with clean samples, though the poisoned samples display a heavier right tail, while temporal triggers produce score profiles that largely coincide with the clean distribution. \textit{In summary, the spike distribution metrics analysis and spike anomaly score distribution analysis demonstrate that temporal triggers resist data-level detection approaches that analyze spike distribution properties, whereas spatiotemporal triggers perturb the input in ways that manifest as clear statistical outliers. The Jitter trigger preserves all content-level and distributional characteristics of the original samples.}


\subsection{Attack Robustness Analysis}


\begin{table*}[!htbp]
\centering
\caption{Performance of various backdoor mitigation methods against spatiotemporal and temporal trigger attacks on neuromorphic datasets. CA: Clean Accuracy (\%), ASR: Attack Success Rate (\%).}
\label{tab:defense_results}
\resizebox{\textwidth}{!}{%
\begin{tabular}{ll|cccccc|cccccc|cccccc}
\toprule
 &  & \multicolumn{6}{c|}{\textbf{N-MNIST}} & \multicolumn{6}{c|}{\textbf{CIFAR10-DVS}} & \multicolumn{6}{c}{\textbf{N-Caltech101}} \\
\textbf{Defense} &  & \textbf{Static} & \textbf{Moving} & \textbf{Smart} & \textbf{Jitter} & \textbf{Latency} & \textbf{Rate} & \textbf{Static} & \textbf{Moving} & \textbf{Smart} & \textbf{Jitter} & \textbf{Latency} & \textbf{Rate} & \textbf{Static} & \textbf{Moving} & \textbf{Smart} & \textbf{Jitter} & \textbf{Latency} & \textbf{Rate} \\
\midrule
\multirow{2}{*}{No Defense} & \textbf{CA} & 99.25 & 99.32 & 99.29 & 99.32 & 99.34 & 99.34 & 70.30 & 69.10 & 70.40 & 71.40 & 67.70 & 71.20 & 73.03 & 77.04 & 75.94 & 74.61 & 75.94 & 77.04 \\
 & \textbf{ASR} & 100.00 & 100.00 & 99.71 & 100.00 & 99.97 & 100.00 & 100.00 & 100.00 & 100.00 & 98.78 & 99.89 & 100.00 & 98.42 & 99.15 & 92.83 & 99.87 & 100.00 & 100.00 \\ \midrule
\multirow{2}{*}{ANP ($thr$ = 0.2)} & \textbf{CA} & 99.21 & 99.22 & 99.20 & 99.15 & 99.10 & 99.26 & 53.40 & 41.00 & 36.40 & 36.70 & 10.00 & 28.60 & 9.60 & 10.45 & 5.59 & 5.59 & 4.98 & 5.59 \\
 & \textbf{ASR} & 100.00 & 100.00 & 99.84 & 100.00 & 0.10 & 57.90 & 99.10 & 81.40 & 62.30 & 99.00 & 100.00 & 3.33 & 0.00 & 14.95 & 100.00 & 100.00 & 99.49 & 99.87 \\ \midrule
\multirow{2}{*}{ANP ($thr$ = 0.5)} & \textbf{CA} & 99.23 & 99.21 & 99.15 & 99.16 & 99.25 & 97.95 & 53.40 & 47.30 & 36.40 & 38.20 & 10.00 & 37.70 & 9.60 & 4.37 & 6.32 & 5.35 & 4.86 & 9.72 \\
 & \textbf{ASR} & 100.00 & 100.00 & 73.56 & 100.00 & 89.76 & 100.00 & 99.10 & 87.50 & 62.30 & 96.78 & 100 & 96.78 & 0.00 & 0.00 & 10.33 & 98.46 & 93.95 & 1.80 \\ \midrule
\multirow{2}{*}{CLP} & \textbf{CA} & 99.18 & 99.29 & 99.32 & 99.33 & 99.34 & 99.32 & 70.80 & 66.00 & 62.50 & 42.80 & 64.40 & 63.30 & 68.29 & 65.98 & 53.71 & 71.93 & 71.93 & 54.68 \\
 & \textbf{ASR} & 100.00 & 100.00 & 97.76 & 100.00 & 77.89 & 100.00 & 100.00 & 100.00 & 100.00 & 100.00 & 96.78 & 99.67 & 98.66 & 99.88 & 99.76 & 43.37 & 100.00 & 95.11 \\ \midrule
\multirow{2}{*}{Fine-Prune} & \textbf{CA} & 99.08 & 99.26 & 99.05 & 99.17 & 99.18 & 99.24 & 69.20 & 66.20 & 67.70 & 18.00 & 25.70 & 66.20 & 71.69 & 71.57 & 72.66 & 70.23 & 63.43 & 72.17 \\
 & \textbf{ASR} & 100.00 & 100.00 & 10.92 & 100.00 & 98.18 & 99.19 & 100.00 & 100.00 & 99.90 & 98.56 & 100.00 & 100.00 & 99.64 & 95.26 & 94.17 & 23.04 & 45.56 & 83.78 \\ \midrule
\multirow{2}{*}{Fine-Tune} & \textbf{CA} & 99.25 & 99.29 & 99.32 & 99.29 & 99.33 & 99.33 & 70.40 & 68.80 & 70.20 & 69.80 & 67.30 & 70.30 & 71.69 & 73.75 & 76.79 & 73.39 & 66.10 & 76.18 \\
 & \textbf{ASR} & 100.00 & 100.00 & 98.72 & 100.00 & 99.99 & 100.00 & 100.00 & 100.00 & 100.00 & 98.67 & 99.89 & 100.00 & 98.42 & 98.30 & 30.26 & 99.87 & 26.13 & 100.00 \\ \midrule
\multirow{2}{*}{NAD} & \textbf{CA} & 99.25 & 99.24 & 99.27 & 99.30 & 99.28 & 99.25 & 69.60 & 69.80 & 70.10 & 70.80 & 67.20 & 71.10 & 26.73 & 2.55 & 6.32 & 5.47 & 15.55 & 18.83 \\
 & \textbf{ASR} & 100.00 & 100.00 & 89.97 & 100.00 & 99.99 & 100.00 & 100.00 & 100.00 & 100.00 & 98.33 & 98.67 & 100.00 & 99.88 & 26.37 & 10.09 & 97.94 & 60.10 & 48.91 \\ \midrule
\multirow{2}{*}{TSBD} & \textbf{CA} & 99.25 & 99.15 & 99.37 & 99.29 & 99.29 & 99.24 & 70.10 & 68.10 & 70.50 & 70.50 & 66.10 & 70.50 & 69.38 & 61.12 & 69.74 & 67.80 & 56.38 & 69.62 \\
 & \textbf{ASR} & 100.00 & 100.00 & 98.84 & 100.00 & 99.75 & 100.00 & 100.00 & 99.80 & 100.00 & 92.00 & 98.89 & 99.78 & 99.51 & 32.44 & 86.76 & 14.41 & 35.91 & 88.42 \\
\bottomrule
\end{tabular}%
}
\end{table*}


\noindent To evaluate the robustness of our proposed T-Backdoor, we utilize nine representative defense methods. We analyze the recently proposed Temporal Membrane Potential Backdoor Detection (TMPBD)~\cite{li2025tmpbd}, two state-of-the-arts backdoor detection methods neural cleasne (NC) \cite{wang2019neural} and STRIP \cite{gao2019strip}, and six model-level mitigation techniques: vanilla fine-tuning, fine-pruning~\cite{liu2018finepruning}, neural attention distillation (NAD)~\cite{li2021nad}, adversarial neuron pruning (ANP)~\cite{wu2021adversarial}, channel lipschitz pruning (CLP)~\cite{zheng2022data}, and two-stage backdoor defense (TSBD)~\cite{lin2024unveiling}. To implement all of the defense methods, we utilize the official published code repositories. The results are presented in Figure~\ref{fig:tmpbd}, \ref{fig:neural_cleanse}, \ref{fig:strip}, and Table~\ref{tab:defense_results}.

\subsubsection{Backdoor Detection Results}

\begin{figure}[!htbp]
\centering
\includegraphics[width=\linewidth]{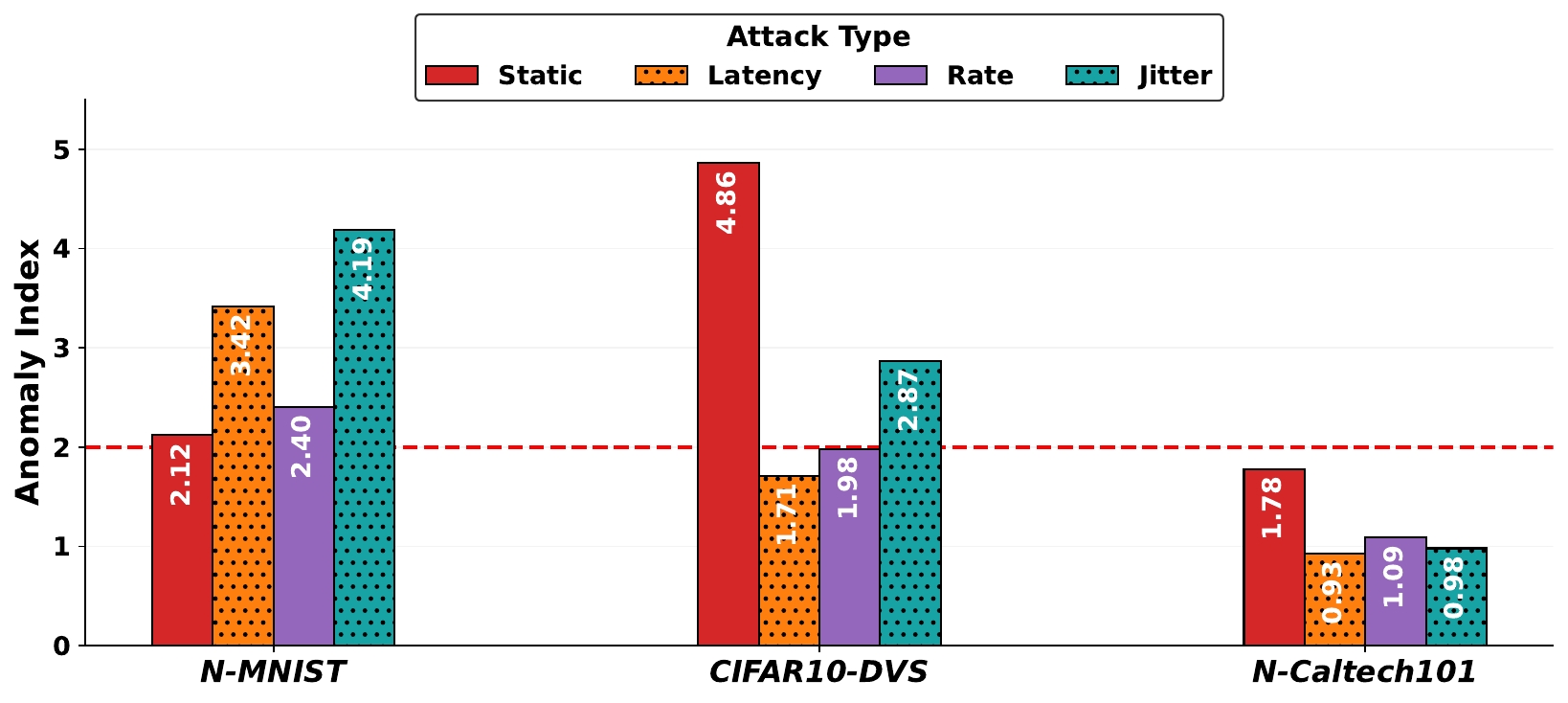}
\caption{Backdoor detection using neural cleanse. The red dot marks the anomaly index; models with a label that exhibits an anomaly index higher than 2 are flagged as backdoored.}
\label{fig:neural_cleanse}
\end{figure}

\begin{figure*}[!htbp]
\centering
\begin{subfigure}[b]{\textwidth}
    \centering
    \includegraphics[width=0.9\textwidth]{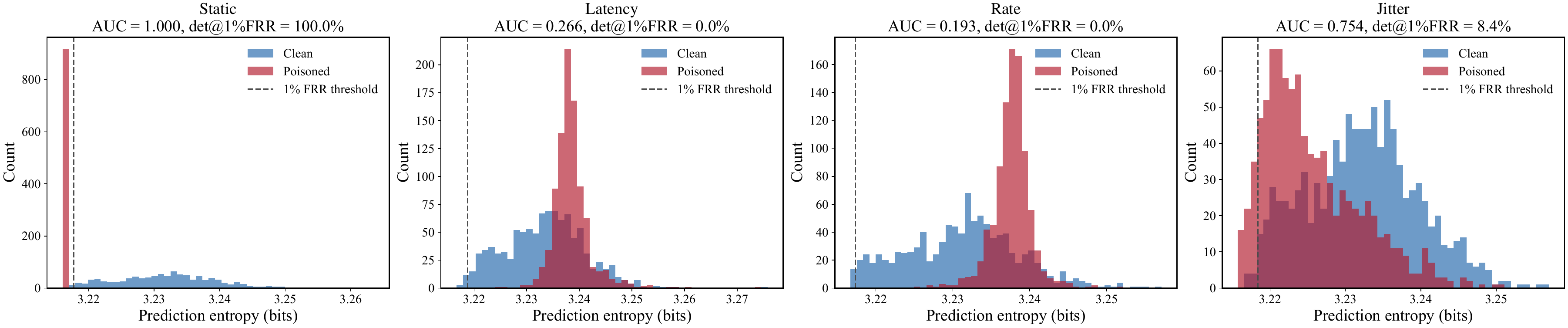}
    \caption{N-MNIST}
    \label{fig:strip_mnist}
\end{subfigure}
\vspace{0.5em}
\begin{subfigure}[b]{\textwidth}
    \centering
    \includegraphics[width=0.9\textwidth]{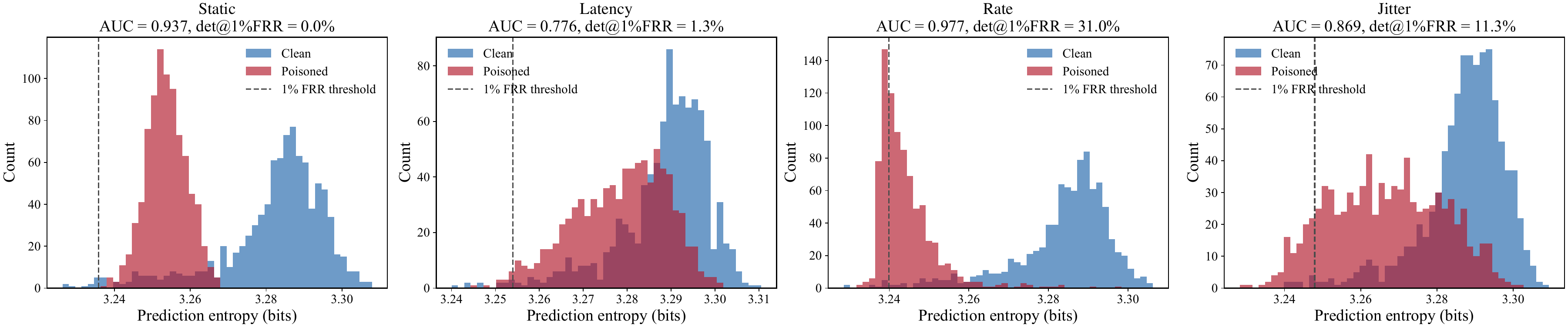}
    \caption{CIFAR10-DVS}
    \label{fig:strip_cifar10}
\end{subfigure}
\vspace{0.5em}
\begin{subfigure}[b]{\textwidth}
    \centering
    \includegraphics[width=0.9\textwidth]{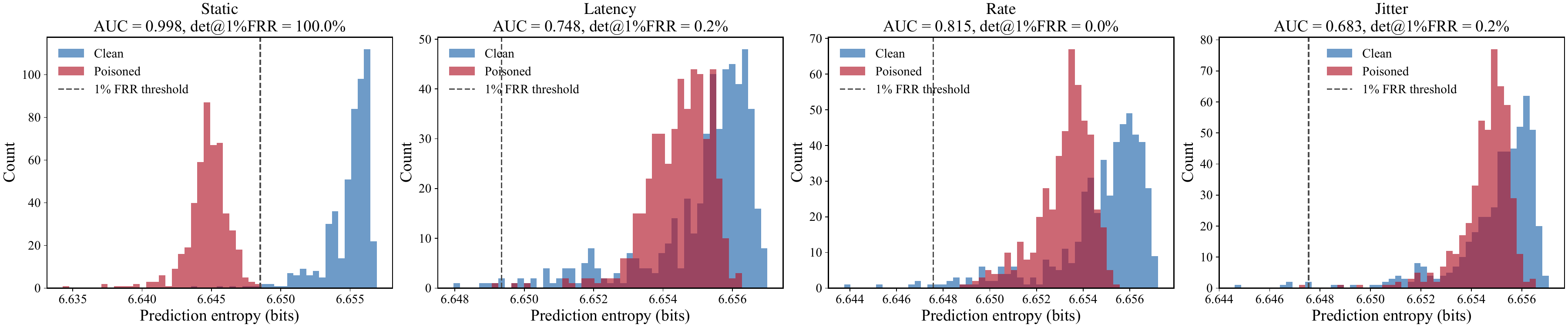}
    \caption{N-Caltech101}
    \label{fig:strip_caltech}
\end{subfigure}
\caption{STRIP prediction entropy distributions for clean and poisoned samples across trigger types, evaluated on N-MNIST, CIFAR10-DVS, and N-Caltech101. AUC: area under ROC curve, det@1\%FRR: detection rate at 1\% false rejection rate (\%).}
\label{fig:strip}
\end{figure*}

We evaluate the TMPBD detection method, which analyzes the temporal membrane potential in the last spiking layer of the SNN to detect the target label based on the intuition that, in the presence of a backdoor, the target label is associated with abnormal overfitting. Such detection method calculates $p$-values via a Gamma-distribution outlier test on a backdoored model, and flags it as tainted if the $p$-value falls below a predefined threshold of $0.05$ as shown in the red dashed line in the Figure~\ref{fig:tmpbd}. From the results, we can see that TMPBD is mainly successful in detecting attacks on the CIFAR10-DVS dataset, while only three attacks are detected on N-MNIST, and none are detected on N-Caltech101. We can observe that detection performance degrades sharply for both spatiotemporal and temporal triggers as dataset and model complexity increase, as evidenced by the highest $p$-values appearing for N-Caltech101. Neural cleanse (NC) proposes a reverse-engineering approach to find the minimum trigger required to misclassify a sample from its true label, along with its $L_1$ norm. Then, from the $L_1$ norms of each label, it proposes an outlier detection algorithm that calculates an anomaly index for every label. If any label exhibits a high anomaly index, it is flagged as the target label, and the model is flagged as a backdoored model. This method was originally developed for conventional image data, and we adapt it for neuromorphic data. The anomaly index calculated by NC for the three temporal triggers and one spatiotemporal trigger (Static), for each dataset, is shown in Figure \ref{fig:neural_cleanse}. From the results, we can see that NC is mainly successful in detecting backdoors for the N-MNIST dataset, but fails to detect 2 attacks on the CIFAR10-DVS dataset and all attacks on the N-Caltech101 dataset. STRIP is mainly proposed for detecting poisoned samples. The detection algorithm is proposed with the intuition that poisoned samples show higher robustness in the presence of strong perturbations than clean samples. So, poisoned samples exhibit low entropy under stronger perturbation. We apply the STRIP algorithm to neuromorphic backdoors for both spatiotemporal and temporal triggers, and the results are presented in Figure \ref{fig:strip}. From the results, it can be seen that poisoned samples with static trigger can be easily detected by STRIP, with AUC close to 1.0 and FRR close to 100\% when the detection threshold is set so that only 1\% of clean samples are mistakenly rejected. However, for temporal triggers like rate, latency, and jitter, the separation between the clean and poisoned sample entropies is limited, and reliable detection is not possible.



\begin{figure}
\centering
\includegraphics[width=\linewidth]{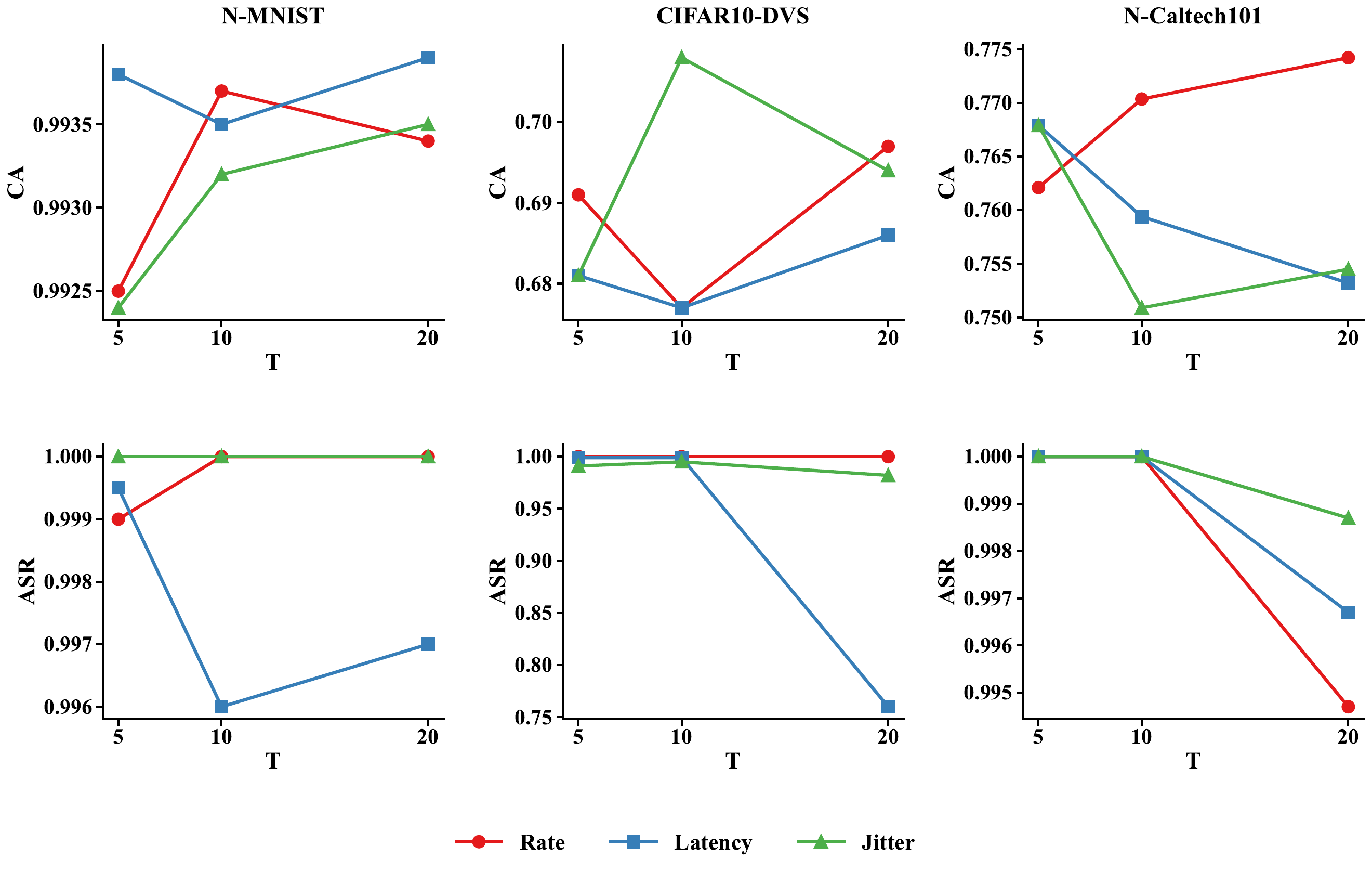}
\caption{Scalability of T-Backdoor over different time steps.}
\label{fig:T_experiment}
\end{figure}

\subsubsection{Backdoor Mitigation Results}

We evaluate the backdoor mitigation techniques that are originally designed for vision models, and adapt them for SNNs presented in Table~\ref{tab:defense_results}. We test with two pruning thresholds for ANP, $thr = 0.2$ and $thr = 0.5$ where more neurons are pruned with a lower $thr$. On N-MNIST, nearly every attack and defense combination leaves ASR at a meaningful level, with only two exceptions: ANP ($thr{=}0.2$) reduces Latency trigger ASR to 0.10\% while preserving 99.10\% CA, and Fine-Pruning reduces Smart ASR to 10.92\% at 99.05\% CA. However, such success does not generalize to other attack types on N-MNIST. Fine-Tuning, NAD, CLP and TSBD are completely ineffective on N-MNIST, indicating that the simple architecture and high baseline accuracy allow backdoor features to survive mild model-level modifications. On CIFAR10-DVS, defense behavior becomes more varied but remains largely unsuccessful. Among 42 configurations, 37 retain ASR above 90\% after applying defense. ANP at $thr {=}0.2$ partially reduces Smart trigger to 62.30\%, Moving to 81.40\%, and Rate to 3.33\%, but at a severe cost to CA with the drops to 10--53\%. ANP completely fails against the Latency trigger, where CA drops to 10\% while ASR remains at 100\%. A similar trend can be seen with Fine-Pruning, which degrades the CA on Jitter and Latency without any meaningful ASR reduction. CLP, Fine-Tuning, NAD and TSBD are essentially ineffective on the dataset, suggesting that the backdoor features on CIFAR10-DVS are deeply entangled with clean ones and cannot be separated by standard distillation or fine-tuning approaches. In contrast, N-Caltech101 presents the most nuanced results. ANP degrades the CA of all attack types below 10\%, rendering any ASR reduction useless. NAD exhibits a similar pattern, lowering CA to 2.55--26.73\%; while it achieves some ASR reduction on Smart (10.09\%) and Rate (48.91\%), the accompanying CA destruction makes these results impractical. TSBD achieves the most consistent partial mitigation, reducing ASR to 14.41\% on Jitter, 32.44\% on Moving, and 35.91\% on Latency while retaining CA in the 56--69\% range. Fine-Pruning provides moderate mitigation on Jitter and Latency. Fine-Tuning presents partial mitigation for Smart and Latency triggers yet completely fails against others. CLP partially mitigates Jitter while remaining ineffective against other attacks. \textit{In summary, T-Backdoor is resistant to TMPBD, NC, and STRIP detection methods and state-of-the-art backdoor mitigation techniques. All the analyzed detection methods fail to detect T-Backdoor in several scenarios, particularly on N-Caltech101. As for the mitigation defenses, 47 out of 63 temporal trigger configurations retain ASR above 90\%, while only 7 could be meaningfully mitigated. These results also demonstrate that a new line of defense needs to be developed that specifically considers the temporal modification of neuromorphic data.}

\begin{table}
\centering
\caption{Single-trigger multi-targeted attacks at $p=0.2$. $N$ = number of targets. Best parameter config per trigger used. \textbf{Bold} = Avg ASR $\geq$ 95\%.}
\label{tab:single_trigger_multi_target}
\begin{tabular}{@{}ll cc cc cc@{}}
\toprule
& & \multicolumn{2}{c}{\textbf{N-MNIST}} & \multicolumn{2}{c}{\textbf{N-Caltech101}} & \multicolumn{2}{c}{\textbf{CIFAR10-DVS}} \\
\cmidrule(lr){3-4} \cmidrule(lr){5-6} \cmidrule(lr){7-8}
\textbf{Trigger} & $N$ & CA & Avg & CA & Avg & CA & Avg \\
\midrule
\multirow{3}{*}{Rate}
  & 3 & 99.23 & \textbf{100.0} & 74.81 & \textbf{99.76} & 68.38 & \textbf{98.89} \\
  & 4 & 99.21 & \textbf{99.99} & 72.78 & \textbf{97.11} & 68.83 & 89.29 \\
  & 5 & 97.88 & \textbf{97.98} & 71.57 & 78.93          & 68.07 & 65.79 \\
\midrule
\multirow{3}{*}{Latency}
  & 3 & 99.29 & \textbf{99.99} & 72.82 & \textbf{99.96} & 66.97 & \textbf{99.71} \\
  & 4 & 99.24 & \textbf{99.98} & 74.48 & \textbf{99.88} & 67.20 & \textbf{99.18} \\
  & 5 & 99.31 & \textbf{99.96} & 76.31 & \textbf{99.71} & 66.90 & \textbf{99.52} \\
\midrule
\multirow{3}{*}{Jitter}
  & 3 & 99.26 & \textbf{100.0} & 72.01 & 66.33 & 67.60 & 94.91 \\
  & 4 & 99.24 & 87.67           & 70.23 & 49.61 & 68.10 & 67.85 \\
  & 5 & 99.32 & 67.22           & 65.74 & 68.09 & 68.20 & 58.18 \\
\bottomrule
\end{tabular}
\vspace{-10pt}
\end{table}


\subsection{Effect of No. of Timesteps $T$ on T-Backdoor}

We investigate whether our proposed T-Backdoor can be generalized across various numbers of timesteps $T \in \{5, 10, 20\}$. In this experiment, the trigger parameters are set as follows: Rate trigger $r{=}0.5$, Latency trigger $d{=}1$, and Jitter $n_s{=}1$. We use a poisoning ratio of $p = 0.1$ for all datasets. As presented in Fig.~\ref{fig:T_experiment}, CA remains mostly stable for N-MNIST, shows mild fluctuation on CIFAR10-DVS, and exhibits a marginal upward trend on N-Caltech for larger $T$, possibly due to improved temporal resolution aiding clean performance on more complex data. On the other hand, ASR is highly robust across all datasets and $T$ values, with the exception on the Latency attack on CIFAR10-DVS at $T{=}20$, where ASR drops to $0.760$. Lower values of $d$ for Latency correspond to a single-timestep shift, which constitutes a smaller perturbation to the spike sequence, reducing the trigger’s distinguishability and lowering its effectiveness. \textit{In summary, The timestep parameter $T$ has minimal impact on CA and generally preserves near-perfect ASR, confirming the robustness of T-Backdoor against varying temporal resolutions.}


\subsection{Comparison with BadSNN}

In this experiment, we compare T-Backdoor with a recently proposed backdoor attack on spiking neural networks named BadSNN \cite{miah2026badsnn}. In this attack, the adversary manipulates the hyperparameters of the LIF neurons of the SNN to embed backdoor behavior into it. BadSNN is proposed for both neuromorphic data and static images. Since T-Backdoor is proposed only for neuromorphic data, we compare it using neuromorphic datasets such as N-MNIST, CIFAR10-DVS, and N-Caltech101. The results are shown in Table \ref{tab:badsnn}. From the results, we can see that all three temporal triggers, Rate, Latency, and Jitter, along with the spatiotemporal trigger, achieve high CA and ASR across all three datasets, whereas BadSNN struggles with ASR on CIFAR10-DVS and N-Caltech101. These results demonstrate the superior performance of T-Backdoor over BadSNN.

\begin{table}[!htbp]
\centering
\caption{Comparison with BadSNN. CA: Clean Accuracy (\%), ASR: Attack Success Rate (\%).}
\label{tab:badsnn}
\begin{tabular}{ll|cc}
\toprule
\textbf{Dataset} & \textbf{Attack} & \textbf{CA} & \textbf{ASR} \\
\midrule
\multirow{6}{*}{N-MNIST} & Static  & 99.25 & 100.00 \\
 & Rate    & 99.34 & 100.00 \\
 & Latency & 99.34 & 99.97  \\
 & Jitter  & 99.32 & 100.00 \\
 & BadSNN & 94.06  & 100.00   \\
\midrule
\multirow{6}{*}{CIFAR10-DVS} & Static  & 70.30 & 100.00 \\
 & Rate    & 71.20 & 100.00 \\
 & Latency & 67.70 & 99.89  \\
 & Jitter  & 71.40 & 98.78  \\
 & BadSNN & 62.66  & 8.44  \\
\midrule
\multirow{6}{*}{N-Caltech101} & Static  & 73.03 & 98.42  \\
 & Rate    & 77.04 & 100.00 \\
 & Latency & 75.94 & 100.00 \\
 & Jitter  & 74.61 & 99.87  \\
 & BadSNN & 64.00  & 0.44  \\
\bottomrule
\end{tabular}
\end{table}

\subsection{Impact of Poisoning Ratio on Attack Effectiveness}

In this experiment, we investigate the effectiveness of the proposed backdoor attack under varying poisoning ratios. Specifically, we vary the poisoning ratio from 1\% to 20\% on the CIFAR10-DVS dataset. The results are presented in Table \ref{tab:cifar10_poison_ratio}. In this experiment, for the rate trigger the scaling factor is set to 0.1x, for the latency trigger the delay frame is set to 1, and for the jitter trigger $n_s = 1$. From the results, we can see that for the rate trigger, the ASR can reach 100\% at even a 1\% poisoning ratio. For latency, however, a 2\% poisoning ratio is needed to reach over 98\% ASR. And for jitter, around 98\% ASR can be reached at a 5\% poisoning ratio. Nonetheless, at a minimum poisoning ratio of 2\%, we observe noticeable attack effectiveness across all three types of backdoor triggers.

\begin{table}[!htbp]
\centering
\caption{Clean accuracy (CA) and attack success rate (ASR) on CIFAR10-DVS across poisoning ratios for different temporal triggers.}
\label{tab:cifar10_poison_ratio}
\begin{tabular}{l|c|cc}
\toprule
\textbf{Trigger} & \textbf{Poisoning Ratio} & \textbf{CA} & \textbf{ASR} \\
\midrule
\multirow{6}{*}{Rate} & 0.01 & 70.3 & 100.0 \\
 & 0.02 & 71.0 & 100.0 \\
 & 0.05 & 71.3 & 100.0 \\
 & 0.10 & 71.3 & 100.0 \\
 & 0.15 & 70.1 & 100.0 \\
 & 0.20 & 70.3 & 100.0 \\
\midrule
\multirow{6}{*}{Latency} & 0.01 & 71.6 & 75.7  \\
 & 0.02 & 70.6 & 98.5  \\
 & 0.05 & 70.5 & 100.0 \\
 & 0.10 & 69.0 & 100.0 \\
 & 0.15 & 67.3 & 100.0 \\
 & 0.20 & 67.9 & 100.0 \\
\midrule
\multirow{6}{*}{Jitter} & 0.01 & 71.5 & 52.2  \\
 & 0.02 & 72.1 & 87.9  \\
 & 0.05 & 72.2 & 98.7  \\
 & 0.10 & 71.0 & 99.9  \\
 & 0.15 & 69.0 & 100.0 \\
 & 0.20 & 71.0 & 100.0 \\
\bottomrule
\end{tabular}
\end{table}

\subsection{Attack Effectiveness Analysis on Source-specific Setting}

In this experiment, we analyze the attack effectiveness in the source-specific attack setting, where samples from source classes are predicted as the wrong (target) label only when triggered, while samples from non-source classes are predicted as the clean label when triggered. The backdoor training approach for this setting is already presented in Section \ref{subsec:backdoor_training}, and the results are presented in Table \ref{tab:source_specific}. From the results, we can see that with three source classes, we can achieve close to or more than 80\% ASR for all three temporal trigger types on the CIFAR10-DVS dataset. However, we can also observe an FTR around 30\% for latency and jitter, and around 41\% for rate. These results show that further modification is necessary in the backdoor training approach to achieve strong attack performance in the source-specific attack setting.

\begin{table}[!htbp]
\centering
\caption{Source-specific backdoor attack performance across trigger types on CIFAR10-DVS dataset. CA: Clean Accuracy (\%), CA-Src: Clean Accuracy on source classes (\%), ASR: Attack Success Rate (\%), FTR: False Trigger Rate (\%).}
\label{tab:source_specific}
\begin{tabular}{l|c|cccc}
\toprule
\textbf{Trigger} & \textbf{Source Classes} & \textbf{CA} & \textbf{CA-Src} & \textbf{ASR} & \textbf{FTR} \\
\midrule
Latency & 1 2 3 & 67.1 & 49.67 & 83.67 & 28.00 \\
Rate    & 1 2 3 & 68.2 & 54.00 & 79.00 & 41.33 \\
Jitter  & 1 2 3 & 67.4 & 51.67 & 80.67 & 26.83 \\
\bottomrule
\end{tabular}
\end{table}

\subsection{Single-Trigger Multi-Target Attack Results Analysis}


\begin{table*}[!htbp]
\centering
\caption{Single-trigger multi-targeted attack results with $N=3$ target classes.
We report the best Avg ASR (\%) and corresponding CA (\%) for each temporal trigger type across datasets at varying poisoning ratio $p$.
\textbf{Bold} = best Avg ASR per dataset.}
\label{tab:main_results}
\begin{adjustbox}{max width=\textwidth}
\renewcommand{\arraystretch}{1.15}
\begin{tabular}{ll ccccc ccccc ccccc}
\toprule
& & \multicolumn{5}{c}{\textbf{N-MNIST}} & \multicolumn{5}{c}{\textbf{N-Caltech101}} & \multicolumn{5}{c}{\textbf{CIFAR10-DVS}} \\
\cmidrule(lr){3-7} \cmidrule(lr){8-12} \cmidrule(lr){13-17}
\textbf{Trigger} & \textbf{Metric} & $p$=0.1 & $p$=0.2 & $p$=0.3 & $p$=0.4 & $p$=0.5
                                    & $p$=0.1 & $p$=0.2 & $p$=0.3 & $p$=0.4 & $p$=0.5
                                    & $p$=0.1 & $p$=0.2 & $p$=0.3 & $p$=0.4 & $p$=0.5 \\
\midrule
\multirow{2}{*}{Rate}
  & CA     & 99.42 & 99.23 & 99.23 & 99.26 & 99.08 & 75.82 & 74.81 & 70.72 & 70.35 & 65.61 & 69.90 & 68.38 & 67.30 & 65.55 & 61.80 \\
  & Avg ASR    & \textbf{100.0} & \textbf{100.0} & \textbf{100.0} & \textbf{100.0} & \textbf{100.0} & 98.99 & \textbf{99.76} & 99.80 & \textbf{99.92} & \textbf{99.96} & 95.70 & 98.89 & \textbf{99.82} & \textbf{99.95} & \textbf{99.77} \\
\midrule
\multirow{2}{*}{Latency}
  & CA     & 99.36 & 99.29 & 99.33 & 99.16 & 99.07 & 75.09 & 72.82 & 71.81 & 69.02 & 65.01 & 68.40 & 66.97 & 65.20 & 63.40 & 59.50 \\
  & Avg ASR    & 99.95 & 99.99 & 99.99 & \textbf{100.0} & 99.99 & \textbf{99.64} & 99.96 & \textbf{100.0} & \textbf{100.0} & 99.96 & \textbf{98.87} & \textbf{99.71} & 99.90 & 99.87 & 99.93 \\
\midrule
\multirow{2}{*}{Jitter}
  & CA     & 99.29 & 99.26 & 99.15 & 99.19 & 99.03 & 75.94 & 72.01 & 70.35 & 64.64 & 56.26 & 69.20 & 67.60 & 65.30 & 64.70 & 60.40 \\
  & Avg ASR    & 99.99 & \textbf{100.0} & \textbf{100.0} & \textbf{100.0} & \textbf{100.0} & 65.78 & 66.33 & 66.18 & 66.46 & 66.50 & 73.83 & 94.91 & 98.90 & 98.93 & 99.03 \\
\bottomrule
\end{tabular}
\end{adjustbox}
\end{table*}


In this experiment, we extend the payload capability of the proposed T-Backdoor to multi-targeted settings. Specifically, we analyze the single-trigger multi-targeted (STMT) scenario, where a single trigger type is parameterized to simultaneously attack $N$ distinct target classes. Each target $k \in \{0, \ldots, N{-}1\}$ is assigned parameter $q_k = k + 1$ for Latency and Jitter. Parameters are evenly spaced over $[0.1,\ r_{\max}]$ for Rate, with $r_{\max} = 4.0$ for $N \in \{3, 4\}$ and $r_{\max} = 3.0$ for $N = 5$. The single-trigger multi-targeted attack results with poisoning ratio $p = 0.2$ for various numbers of targets $N$ are given in Table~\ref{tab:single_trigger_multi_target}. It can be seen that the Latency trigger performs best, maintaining ASR $\geq$ 99\% up to $N=5$ on all three datasets.

\begin{figure}
    \centering
    \includegraphics[width=\linewidth]{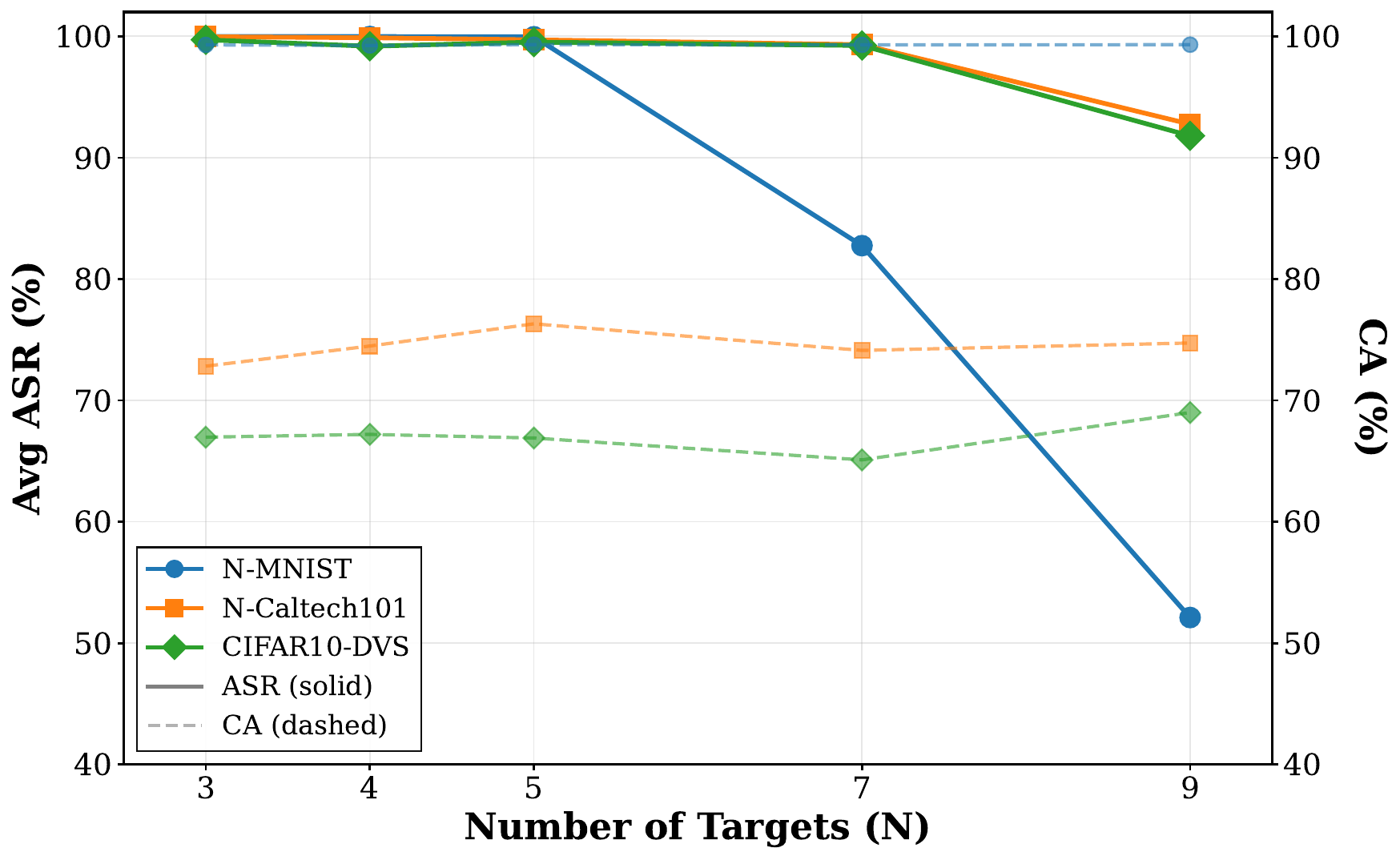}
    \caption{Latency trigger's multi-targeted attack capability.}
    \label{fig:latency_multi_target}
\end{figure}

\noindent We further test Latency's scalability by increasing $N$ up to 9 shown in Figure~\ref{fig:latency_multi_target}. Latency scales effectively to $N=9$ on CIFAR10-DVS and N-Caltech101, but falls short on N-MNIST. Rate trigger maintains strong performance up to $N=4$ but degrades sharply at $N=5$, dropping to 78.93\% on N-Caltech101 and 65.79\% on CIFAR10-DVS. Jitter is the least scalable, losing most of its effectiveness at $N \geq 4$, particularly on N-Caltech101. On the contrary, CA remains stable, indicating that increasing targets does not compound clean accuracy loss beyond what the fixed poisoning rate already introduces. We also vary $p \in \{0.1, 0.2, 0.3, 0.4, 0.5\}$ for a fixed $N=3$ in Table~\ref{tab:main_results}. Both Rate and Latency achieve near-perfect ASRs across all datasets and poisoning ratios, whereas Jitter exhibits dataset-dependent limitations, achieving perfect ASR on N-MNIST and CIFAR10-DVS for $p > 0.2$, but underperforming on N-Caltech101. CA is well preserved on N-MNIST yet degrades progressively with increasing $p$ on N-Caltech101 and CIFAR10-DVS. \textit{In summary, the single-trigger multi-targeted analysis using T-backdoor demonstrates that Latency is the most stable and scalable trigger for multi-targeted scenarios, while Jitter is the least effective. Overall, multi-targeted payloads can be achieved by T-Backdoor with decent performance.}

\subsection{Multi-Trigger Multi-Targeted Attack Analysis}

In this experiment, we evaluate the scalability of the proposed T-Backdoor with a multi-trigger multi-targeted attack analysis, where multiple temporal triggers are simultaneously embedded in the backdoored model with each mapped to a distinct target class. Specifically, when the Rate trigger is used once, $r{=}0.5$ is applied; when used twice, $r{=}0.5$ and $r{=}2.0$ are used for the two respective targets. Similarly, the Latency trigger uses $d{=}1$ for a single instance and $d{=}1$ and $d{=}2$ for two instances. $n_s{=}3$ is used for Jitter for a single instance, while $n_s{=}2$ and $n_s{=}3$ are used for two instances. Note that the clean-model CA baselines are 99.32\% (N-MNIST), 78.37\% (N-Caltech101), and 71.00\% (CIFAR10-DVS). 

\begin{table}
\centering
\caption{MTMT attack results on N-Caltech101 and CIFAR10-DVS.
Each $TR_i$ shows per-target ASR (\%).
$C_1$: L+R+J ($N\!=\!3$),
$C_2$: L+L+R+J ($N\!=\!4$),
$C_3$: L+L+R+R+J ($N\!=\!5$),
$C_4$: L+L+R+J+J ($N\!=\!5$),
$C_5$: L+R+R+J+J ($N\!=\!5$). Here L\,=\,Latency, R\,=\,Rate, J\,=\,Jitter.
On N-MNIST, all configs achieve 100\% ASR on every target with CA\,$\geq$\,99.37\%.
\textbf{Bold}\,=\,Avg $\geq$\,99\%.}
\label{tab:mtmt}
\begin{tabular}{@{}lcc ccccc c@{}}
\toprule
\textbf{Cfg} & $p$ & \textbf{CA} & $TR_0$ & $TR_1$ & $TR_2$ & $TR_3$ & $TR_4$ & \textbf{Avg} \\
\midrule
\multicolumn{9}{@{}l}{\textit{N-Caltech101}} \\
$C_1$ & .3 & 70.47 & 100.0 & 100.0 & 99.64 & — & — & \textbf{99.88} \\
$C_2$ & .1 & 75.94 & 99.88 & 99.64 & 99.88 & 97.81 & — & \textbf{99.30} \\
$C_2$ & .2 & 73.15 & 100.0 & 100.0 & 99.51 & 99.64 & — & \textbf{99.79} \\
$C_2$ & .3 & 72.54 & 100.0 & 99.88 & 100.0 & 99.51 & — & \textbf{99.85} \\
$C_3$ & .1 & 73.51 & 99.39 & 99.64 & 99.27 & 99.15 & 97.45 & 98.98 \\
$C_3$ & .2 & 71.63 & 99.76 & 99.94 & 99.76 & 99.64 & 99.57 & \textbf{99.73} \\
$C_3$ & .3 & 70.60 & 100.0 & 100.0 & 99.64 & 99.76 & 99.76 & \textbf{99.83} \\
$C_4$ & .2 & 70.11 & 99.88 & 100.0 & 99.64 & 98.30 & 98.78 & \textbf{99.32} \\
$C_5$ & .2 & 68.41 & 100.0 & 100.0 & 99.51 & 97.81 & 99.03 & \textbf{99.27} \\
\midrule
\multicolumn{9}{@{}l}{\textit{CIFAR10-DVS}} \\
$C_1$ & .3 & 63.60 & 100.0 & 100.0 & 98.80 & — & — & \textbf{99.60} \\
$C_2$ & .1 & 69.10 & 93.05 & 98.65 & 99.70 & 73.65 & — & 91.26 \\
$C_2$ & .2 & 66.50 & 90.30 & 100.0 & 98.50 & 95.60 & — & 96.10 \\
$C_2$ & .3 & 63.80 & 77.40 & 100.0 & 100.0 & 97.90 & — & 93.82 \\
$C_3$ & .1 & 68.80 & 87.40 & 98.55 & 99.75 & 96.60 & 52.25 & 86.91 \\
$C_3$ & .2 & 66.75 & 97.55 & 100.0 & 99.95 & 98.25 & 91.00 & 97.35 \\
$C_3$ & .3 & 64.75 & 86.25 & 100.0 & 99.80 & 99.50 & 93.80 & 95.87 \\
$C_4$ & .2 & 67.25 & 87.35 & 100.0 & 99.15 & 82.25 & 86.25 & 91.00 \\
$C_5$ & .2 & 67.20 & 99.65 & 99.90 & 98.90 & 79.00 & 89.45 & 93.38 \\
\bottomrule
\end{tabular}
\end{table}

\noindent As stated in Table~\ref{tab:mtmt}, all trigger configurations of T-Backdoor achieve a perfect 100\% Avg ASR across all configurations and poisoning rates, while the CA remains consistent with the clean-model baseline. The performance remains strong for the more challenging 101-class setting of N-Caltech101, achieving over 98\% Avg ASR across all configurations. Clean accuracy shows a modest decline relative to the 78.37\% baseline: $C_2$ at $p{=}0.1$ retains 75.94\% ($-$2.43\%), but higher poisoning rates and more triggers increase the drop, $C_5$ at $p{=}0.2$ falls to 68.41\% ($-$9.96\%). The CIFAR10-DVS dataset presents the most challenging scenario due to its higher visual complexity and fewer training samples per class. While the Avg ASR in most cases exceeds 90\%, the Avg ASR drops to 86.91\% at $p{=}0.1$ with the $C_3$ trigger configuration, primarily due to $T_4$ achieving only 52.25\%. However, increasing $p$ to 0.2 substantially recovers the performance. The CA drops remain in the acceptable range of 1.90--7.40\%, which are smaller than N-Caltech101. \textit{In summary, the proposed T-Backdoor scales effectively to the MTMT setting, achieving near-perfect Avg ASRs in most trigger configurations with minimal CA degradation.}

\subsection{Impact of Poisoning Ratio on MTMT Attack Setting}

In this experiment, we investigate the impact of the poisoning ratio on attack effectiveness in the MTMT attack setting, to understand the minimum poisoning ratio needed to achieve strong attack performance in this setting. The results are presented in Table \ref{tab:cifar10dvs_multitarget}. This experiment is conducted on the CIFAR10-DVS dataset. Note that the poisoning ratio denoted here is the overall poisoning ratio. From the results, we can see that with only a 10\% poisoning ratio it is possible to reach an average ASR over 86\%, and with a 20\% poisoning ratio the average ASR reaches up to 97.46\%, with only a 1\% drop in CA. 

\begin{table}[!htbp]
\centering
\caption{Multi-Trigger Multi-Targeted Attack backdoor attack performance on CIFAR10-DVS across poisoning ratios ($\epsilon$) and trigger types. ASR: Attack Success Rate (\%), CA: Clean Accuracy (\%).}
\label{tab:cifar10dvs_multitarget}
\resizebox{\linewidth}{!}{%
\begin{tabular}{c|ll|c|cc}
\toprule
\textbf{Poisoning Ratio} & \textbf{Trigger} & \textbf{Trigger Description} & \textbf{ASR} & \textbf{CA} & \textbf{Avg. ASR} \\
\midrule
\multirow{5}{*}{0.05} & Latency & Target 0: Latency ($d$=1) & 34.6 & \multirow{5}{*}{70.6} & \multirow{5}{*}{66.3} \\
 & Latency & Target 1: Latency ($d$=2) & 89.2 & & \\
 & Rate & Target 2: Rate ($r$=0.5x) & 99.5 & & \\
 & Rate & Target 3: Rate ($r$=2.0x) & 96.5 & & \\
 & Jitter & Target 4: Jitter ($n_s$=3) & 18.8 & & \\
\midrule
\multirow{5}{*}{0.1} & Latency & Target 0: Latency ($d$=1) & 84.5 & \multirow{5}{*}{69.5} & \multirow{5}{*}{86.1} \\
 & Latency & Target 1: Latency ($d$=2) & 99.6 & & \\
 & Rate & Target 2: Rate ($r$=0.5x) & 100.0 & & \\
 & Rate & Target 3: Rate ($r$=2.0x) & 99.1 & & \\
 & Jitter & Target 4: Jitter ($n_s$=3) & 62.4 & & \\
\midrule
\multirow{5}{*}{0.2} & Latency & Target 0: Latency ($d$=1) & 99.6 & \multirow{5}{*}{68.4} & \multirow{5}{*}{97.46} \\
 & Latency & Target 1: Latency ($d$=2) & 100.0 & & \\
 & Rate & Target 2: Rate ($r$=0.5x) & 100.0 & & \\
 & Rate & Target 3: Rate ($r$=2.0x) & 99.5 & & \\
 & Jitter & Target 4: Jitter ($n_s$=3) & 95.2 & & \\
\midrule
\multirow{5}{*}{0.3} & Latency & Target 0: Latency ($d$=1) & 99.7 & \multirow{5}{*}{65.1} & \multirow{5}{*}{98.5} \\
 & Latency & Target 1: Latency ($d$=2) & 100.0 & & \\
 & Rate & Target 2: Rate ($r$=0.5x) & 100.0 & & \\
 & Rate & Target 3: Rate ($r$=2.0x) & 99.8 & & \\
 & Jitter & Target 4: Jitter ($n_s$=3) & 99.2 & & \\
\midrule
\multirow{5}{*}{0.4} & Latency & Target 0: Latency ($d$=1) & 99.9 & \multirow{5}{*}{66.0} & \multirow{5}{*}{99.14} \\
 & Latency & Target 1: Latency ($d$=2) & 100.0 & & \\
 & Rate & Target 2: Rate ($r$=0.5x) & 100.0 & & \\
 & Rate & Target 3: Rate ($r$=2.0x) & 99.9 & & \\
 & Jitter & Target 4: Jitter ($n_s$=3) & 98.8 & & \\
\bottomrule
\end{tabular}%
}
\end{table}

\begin{figure*}[!htbp]
\centering
\begin{subfigure}[t]{0.3\linewidth}
  \centering
  \includegraphics[width=\linewidth]{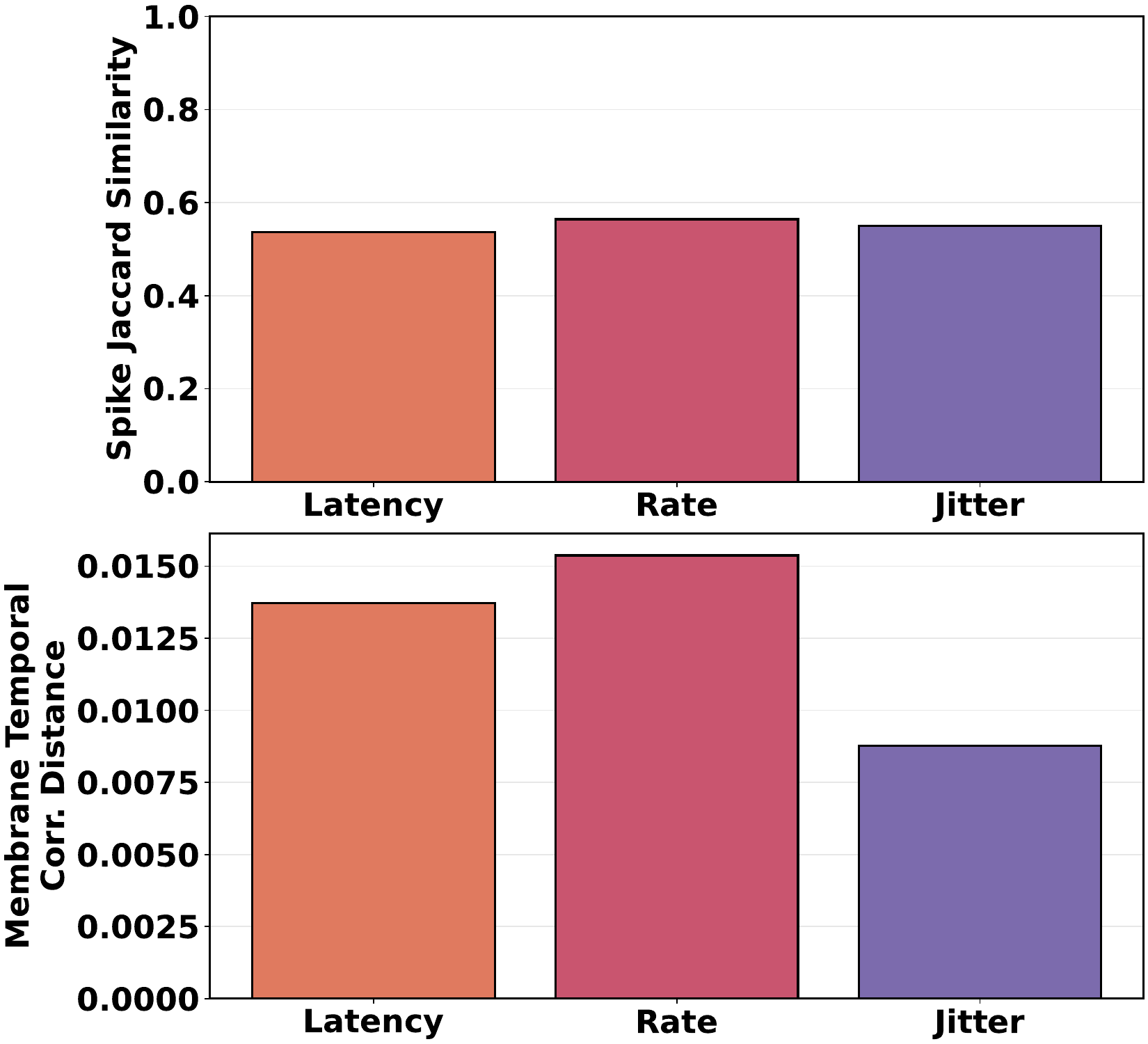}
  \caption{N-MNIST}
  \label{fig:defense_mnist}
\end{subfigure}
\hfill
\begin{subfigure}[t]{0.3\linewidth}
  \centering
  \includegraphics[width=\linewidth]{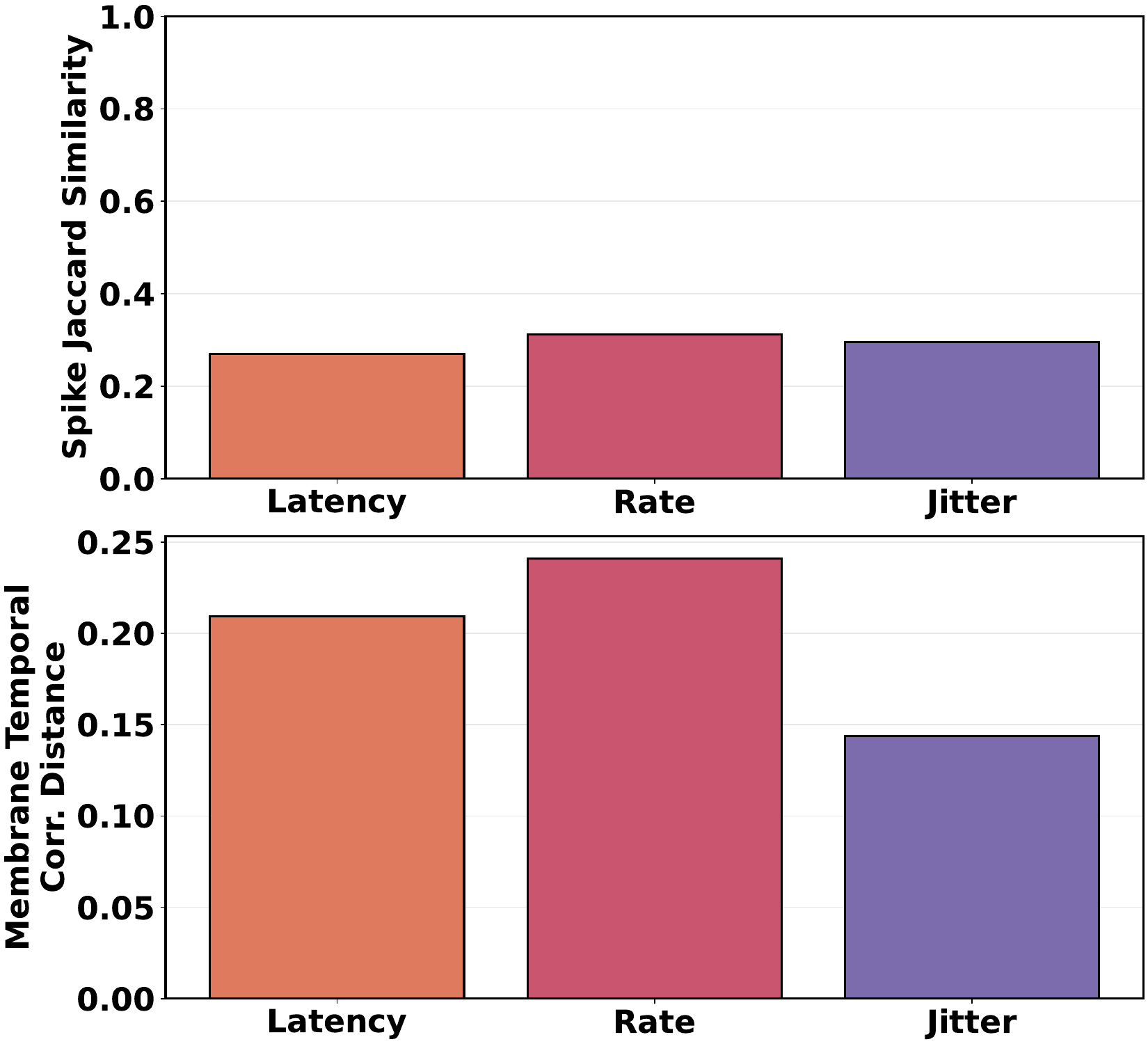}
  \caption{CIFAR10-DVS}
  \label{fig:defense_cifar}
\end{subfigure}
\hfill
\begin{subfigure}[t]{0.3\linewidth}
  \centering
  \includegraphics[width=\linewidth]{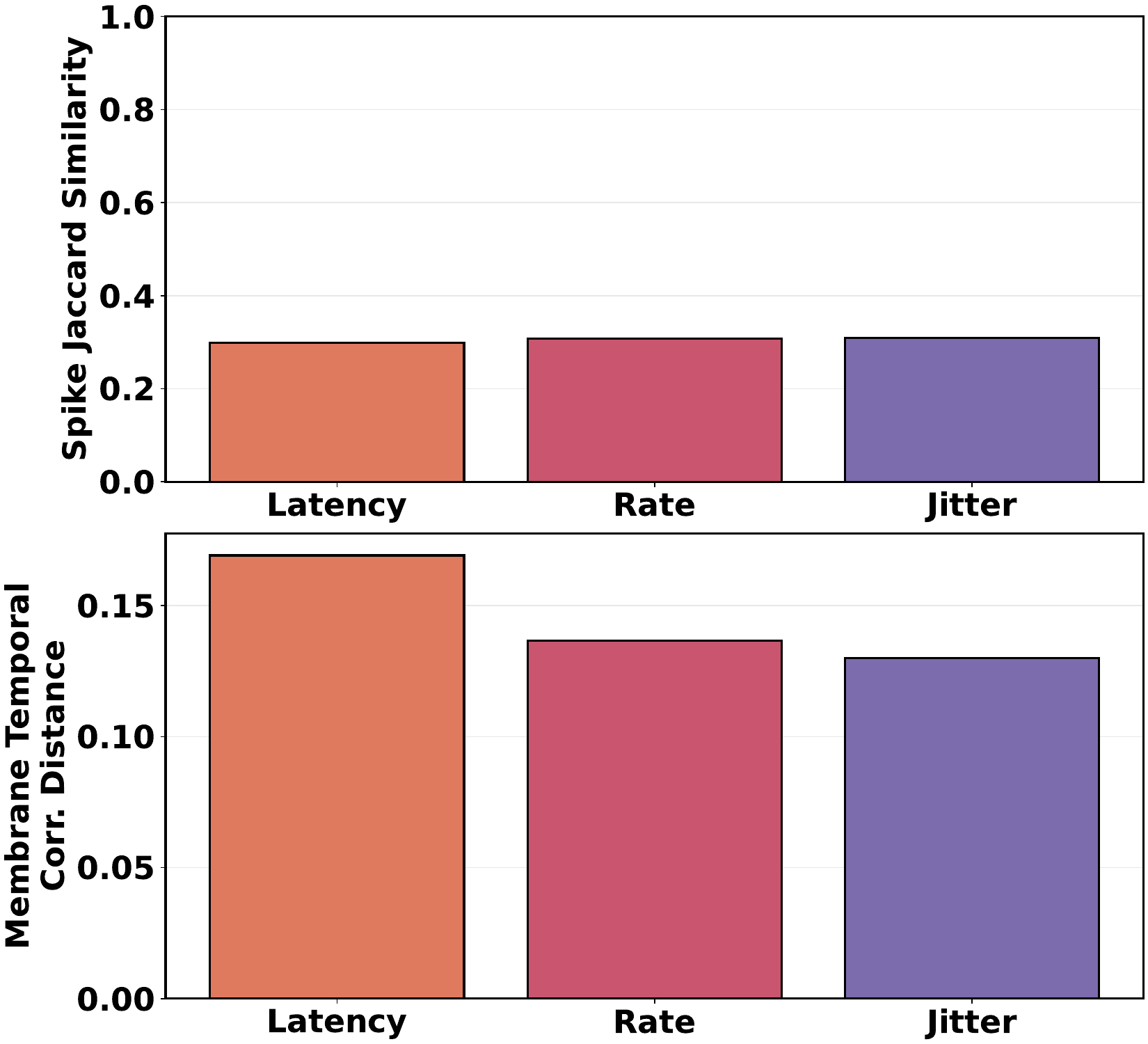}
  \caption{N-Caltech101}
  \label{fig:defense_caltech}
\end{subfigure}
\caption{Model-internal signatures of T-Backdoor activations across
datasets for potential defense analysis. Top: mean spike jaccard similarity $\bar{J}$
between clean and triggered spike trains. Bottom:
membrane temporal correlation distance $D^{\mathrm{MTC}}$. }
\label{fig:defense_signals}
\end{figure*}


\subsection{Discussion and Future Works}
\label{sec:potential_defense_analysis}

\begin{equation}
\begin{aligned}
J_\ell(t) &= \frac{\lVert s^c_\ell(t)\wedge s^p_\ell(t)\rVert_1}
                  {\lVert s^c_\ell(t)\vee s^p_\ell(t)\rVert_1},\quad
\bar{J}_\ell = \frac{1}{T}\sum_{t=0}^{T-1} J_\ell(t), \\
D^{\mathrm{MTC}}_\ell &= \sqrt{\frac{1}{T}\sum_{\tau=0}^{T-1}
     \bigl(R^c_\ell(\tau)-R^p_\ell(\tau)\bigr)^2}.
\end{aligned}
\label{eq:jaccar}
\end{equation}


\noindent While the temporal triggers proposed in T-Backdoor are stealthy in terms of the spike distribution between poisoned and clean counterparts, we investigate two model-level signatures that can be exploited by a defender to detect temporally-triggered poisoned samples. Let $s^c_\ell(t)\in\{0,1\}^{N_\ell}$ and $s^p_\ell(t)\in\{0,1\}^{N_\ell}$ denote the spiking vectors of a spiking layer $\ell$ at timestep $t$ under clean and poisoned samples, respectively, and let $v^c_\ell(t),v^p_\ell(t)\in\mathbb{R}^{N_\ell}$ denote the corresponding membrane-potential vectors. Based on these vectors we define two metrics: (i)~the spike jaccard similarity ($\bar{J}_\ell$) and (ii)~the membrane temporal correlation distance ($D^{\mathrm{MTC}}_\ell$), expressed in the equation \ref{eq:jaccar}, respectively. $\bar{J}_\ell$ captures the overlap of the active-neuron sets between clean and poisoned counterparts: a value of $1.0$ indicates that the same neurons fire on both passes, while lower values indicate that different neurons are activated. $D^{\mathrm{MTC}}_\ell$ measures how much the trigger perturbs the \emph{temporal structure} of the membrane-potential trajectory, independent of its instantaneous value. Figure~\ref{fig:defense_signals} reports both metrics across all three datasets. From the values of $\bar{J}_\ell$ we can see that on the larger datasets CIFAR10-DVS and N-Caltech101, roughly two thirds of the spikes in a poisoned forward pass either fire at a neuron that was silent on the clean pass or fail to fire at a neuron that was active. The $D^{\mathrm{MTC}}_\ell$ values convey a consistent story: although the distance is small on N-MNIST, it grows by more than an order of magnitude on CIFAR10-DVS and N-Caltech101, showing that the temporal triggers substantially disrupt the temporal structure of the membrane-voltage trajectory. Together, these two observables leave a pronounced signature of the temporal triggers in the model's internal dynamics, a signature that a defender can exploit to detect poisoned samples. We will extend a full detector implementation to our future work.

\section{Conclusion} \label{sec:conclusion}


In this paper, we propose T-Backdoor, a spike-preserving temporal backdoor attack on SNNs. Specifically, we introduce three temporal triggers: Rate, Latency and Jitter. Our experiments show that the proposed attack achieves near 100\% ASR in both single-targeted and multi-targeted settings. We believe this work will inspire the community to develop defense methods against backdoor attacks on SNNs, particularly those involving temporal triggers.

\bibliographystyle{unsrt}
\bibliography{acmart}

@inproceedings{gao2019strip,
  title={Strip: A defence against trojan attacks on deep neural networks},
  author={Gao, Yansong and Xu, Change and Wang, Derui and Chen, Shiping and Ranasinghe, Damith C and Nepal, Surya},
  booktitle={Proceedings of the 35th annual computer security applications conference},
  pages={113--125},
  year={2019}
}

@inproceedings{liu2023detecting,
  title={Detecting backdoors during the inference stage based on corruption robustness consistency},
  author={Liu, Xiaogeng and Li, Minghui and Wang, Haoyu and Hu, Shengshan and Ye, Dengpan and Jin, Hai and Wu, Libing and Xiao, Chaowei},
  booktitle={Proceedings of the IEEE/CVF Conference on Computer Vision and Pattern Recognition},
  pages={16363--16372},
  year={2023}
}

@inproceedings{li2025unsupervised,
  title={Unsupervised backdoor detection and mitigation for spiking neural networks},
  author={Li, Jiachen and Wu, Bang and Xia, Xiaoyu and Liu, Xiaoning and Yi, Xun and Zhang, Xiuzhen},
  booktitle={2025 28th International Symposium on Research in Attacks, Intrusions and Defenses (RAID)},
  pages={50--64},
  year={2025},
  organization={IEEE}
}

@inproceedings{zeng2021rethinking,
  title={Rethinking the backdoor attacks' triggers: A frequency perspective},
  author={Zeng, Yi and Park, Won and Mao, Z Morley and Jia, Ruoxi},
  booktitle={Proceedings of the IEEE/CVF international conference on computer vision},
  pages={16473--16481},
  year={2021}
}

@inproceedings{liu2019abs,
  author       = {Yingqi Liu and Wen-Chuan Lee and Guanhong Tao and Shengzhi Ma and Yanjun Aafer and Xiangyu Zhang},
  title        = {{ABS}: Scanning Neural Networks for Back-doors by Artificial Brain Stimulation},
  booktitle    = {Proceedings of the 2019 ACM SIGSAC Conference on Computer and Communications Security},
  year         = {2019},
  pages        = {1265--1282},
  doi          = {10.1145/3319535.3363216},
  publisher    = {Association for Computing Machinery},
  address      = {New York, NY, USA},
  url          = {https://doi.org/10.1145/3319535.3363216}
}

@inproceedings{wang2019neural,
  title={Neural cleanse: Identifying and mitigating backdoor attacks in neural networks},
  author={Wang, Bolun and Yao, Yuanshun and Shan, Shawn and Li, Huiying and Viswanath, Bimal and Zheng, Haitao and Zhao, Ben Y},
  booktitle={2019 IEEE symposium on security and privacy (SP)},
  pages={707--723},
  year={2019},
  organization={IEEE}
}

@inproceedings{zhu2023enhancing,
  title={Enhancing fine-tuning based backdoor defense with sharpness-aware minimization},
  author={Zhu, Mingli and Wei, Shaokui and Shen, Li and Fan, Yanbo and Wu, Baoyuan},
  booktitle={Proceedings of the IEEE/CVF International Conference on Computer Vision},
  pages={4466--4477},
  year={2023}
}

@inproceedings{zheng2022data,
  title={Data-free backdoor removal based on channel lipschitzness},
  author={Zheng, Runkai and Tang, Rongjun and Li, Jianze and Liu, Li},
  booktitle={European Conference on Computer Vision},
  pages={175--191},
  year={2022},
  organization={Springer}
}

@article{wu2021adversarial,
  title={Adversarial neuron pruning purifies backdoored deep models},
  author={Wu, Dongxian and Wang, Yisen},
  journal={Advances in Neural Information Processing Systems},
  volume={34},
  pages={16913--16925},
  year={2021}
}

@inproceedings{li2023reconstructive,
  title={Reconstructive neuron pruning for backdoor defense},
  author={Li, Yige and Lyu, Xixiang and Ma, Xingjun and Koren, Nodens and Lyu, Lingjuan and Li, Bo and Jiang, Yu-Gang},
  booktitle={International Conference on Machine Learning},
  pages={19837--19854},
  year={2023},
  organization={PMLR}
}

@article{miah2026deepcurer,
  title={DeepCurer: Pruning-based backdoor mitigation via progressive neuron ranking using adversarial proxies},
  author={Miah, Abdullah Arafat and Bi, Yu},
  journal={Neurocomputing},
  pages={134409},
  year={2026},
  publisher={Elsevier}
}

@article{gu2019badnets,
  title={Badnets: Evaluating backdooring attacks on deep neural networks},
  author={Gu, Tianyu and Liu, Kang and Dolan-Gavitt, Brendan and Garg, Siddharth},
  journal={IEEE Access},
  volume={7},
  pages={47230--47244},
  year={2019},
  publisher={IEEE}
}

@inproceedings{chen2021badnl,
  title={Badnl: Backdoor attacks against nlp models with semantic-preserving improvements},
  author={Chen, Xiaoyi and Salem, Ahmed and Chen, Dingfan and Backes, Michael and Ma, Shiqing and Shen, Qingni and Wu, Zhonghai and Zhang, Yang},
  booktitle={Proceedings of the 37th Annual Computer Security Applications Conference},
  pages={554--569},
  year={2021}
}

@article{turner2018clean,
  title={Clean-label backdoor attacks},
  author={Turner, Alexander and Tsipras, Dimitris and Madry, Aleksander},
  year={2018}
}

@article{nguyen2021wanet,
  title={Wanet--imperceptible warping-based backdoor attack},
  author={Nguyen, Anh and Tran, Anh},
  journal={arXiv preprint arXiv:2102.10369},
  year={2021}
}

@inproceedings{gao2024dual,
  title={A dual stealthy backdoor: From both spatial and frequency perspectives},
  author={Gao, Yudong and Chen, Honglong and Sun, Peng and Li, Junjian and Zhang, Anqing and Wang, Zhibo and Liu, Weifeng},
  booktitle={Proceedings of the AAAI Conference on Artificial Intelligence},
  volume={38},
  number={3},
  pages={1851--1859},
  year={2024}
}

@inproceedings{pan2022hidden,
  title={Hidden trigger backdoor attack on $\{$NLP$\}$ models via linguistic style manipulation},
  author={Pan, Xudong and Zhang, Mi and Sheng, Beina and Zhu, Jiaming and Yang, Min},
  booktitle={31st USENIX Security Symposium (USENIX Security 22)},
  pages={3611--3628},
  year={2022}
}

@inproceedings{feng2022fiba,
  title={Fiba: Frequency-injection based backdoor attack in medical image analysis},
  author={Feng, Yu and Ma, Benteng and Zhang, Jing and Zhao, Shanshan and Xia, Yong and Tao, Dacheng},
  booktitle={Proceedings of the IEEE/CVF Conference on Computer Vision and Pattern Recognition},
  pages={20876--20885},
  year={2022}
}

@article{chen2017targeted,
  title={Targeted backdoor attacks on deep learning systems using data poisoning},
  author={Chen, Xinyun and Liu, Chang and Li, Bo and Lu, Kimberly and Song, Dawn},
  journal={arXiv preprint arXiv:1712.05526},
  year={2017}
}

@article{doan2021backdoor,
  title={Backdoor attack with imperceptible input and latent modification},
  author={Doan, Khoa and Lao, Yingjie and Li, Ping},
  journal={Advances in Neural Information Processing Systems},
  volume={34},
  pages={18944--18957},
  year={2021}
}

@inproceedings{cheng2021deep,
  title={Deep feature space trojan attack of neural networks by controlled detoxification},
  author={Cheng, Siyuan and Liu, Yingqi and Ma, Shiqing and Zhang, Xiangyu},
  booktitle={Proceedings of the AAAI Conference on Artificial Intelligence},
  volume={35},
  number={2},
  pages={1148--1156},
  year={2021}
}

@inproceedings{wang2022bppattack,
  title={Bppattack: Stealthy and efficient trojan attacks against deep neural networks via image quantization and contrastive adversarial learning},
  author={Wang, Zhenting and Zhai, Juan and Ma, Shiqing},
  booktitle={Proceedings of the IEEE/CVF conference on computer vision and pattern recognition},
  pages={15074--15084},
  year={2022}
}

@article{lin2024unveiling,
  title={Unveiling and mitigating backdoor vulnerabilities based on unlearning weight changes and backdoor activeness},
  author={Lin, Weilin and Liu, Li and Wei, Shaokui and Li, Jianze and Xiong, Hui},
  journal={Advances in Neural Information Processing Systems},
  volume={37},
  pages={42097--42122},
  year={2024}
}

@article{maass1997networks,
  title={Networks of spiking neurons: the third generation of neural network models},
  author={Maass, Wolfgang},
  journal={Neural networks},
  volume={10},
  number={9},
  pages={1659--1671},
  year={1997},
  publisher={Elsevier}
}

@article{neftci2019surrogate,
  title   = {Surrogate Gradient Learning in Spiking Neural Networks: Bringing the Power of Gradient-Based Optimization to Spiking Neural Networks},
  author  = {Neftci, Emre O. and Mostafa, Hesham and Zenke, Friedemann},
  journal = {IEEE Signal Processing Magazine},
  volume  = {36},
  number  = {6},
  pages   = {51--63},
  year    = {2019},
  doi     = {10.1109/MSP.2019.2931595}
}

@article{fang2023spikingjelly,
  title   = {SpikingJelly: An Open-Source Machine Learning Infrastructure Platform for Spike-Based Intelligence},
  author  = {Fang, Wei and Chen, Yanqi and Ding, Jianhao and Yu, Zhaofei and others},
  journal = {Science Advances},
  volume  = {9},
  number  = {42},
  pages   = {eadi1480},
  year    = {2023},
  doi     = {10.1126/sciadv.adi1480}
}

@article{li2017cifar10dvs,
  title   = {CIFAR10-DVS: An Event-Stream Dataset for Object Classification},
  author  = {Li, Hongmin and Liu, Hanchao and Ji, Xiangyang and Li, Guoqi and Shi, Luping},
  journal = {Frontiers in Neuroscience},
  volume  = {11},
  pages   = {309},
  year    = {2017},
  doi     = {10.3389/fnins.2017.00309}
}

@article{davies2018loihi,
  title   = {Loihi: A Neuromorphic Manycore Processor with On-Chip Learning},
  author={Davies, Mike and Srinivasa, Narayan and Lin, Tsung-Han and Chinya, Gautham and Cao, Yongqiang and Choday, Sri Harsha and Dimou, Georgios and Joshi, Prasad and Imam, Nabil and Jain, Shweta and others},
  journal = {IEEE Micro},
  volume  = {38},
  number  = {1},
  pages   = {82--99},
  year    = {2018},
  doi     = {10.1109/MM.2018.112130359}
}

@article{miah2024noiseattack,
  title={Noiseattack: An evasive sample-specific multi-targeted backdoor attack through white gaussian noise},
  author={Miah, Abdullah Arafat and Icer, Kaan and Sendag, Resit and Bi, Yu},
  journal={arXiv preprint arXiv:2409.02251},
  year={2024}
}

@article{miah2026lite,
  title={Lite-BD: A Lightweight Black-box Backdoor Defense via Reviving Multi-Stage Image Transformations},
  author={Miah, Abdullah Arafat and Bi, Yu},
  journal={arXiv preprint arXiv:2602.07197},
  year={2026}
}

@article{khan2026multi,
  title={Multi-Targeted Graph Backdoor Attack},
  author={Khan, Md Nabi Newaz and Miah, Abdullah Arafat and Bi, Yu},
  journal={arXiv preprint arXiv:2601.15474},
  year={2026}
}

@article{miah2024exploiting,
  title={Exploiting the vulnerability of large language models via defense-aware architectural backdoor},
  author={Miah, Abdullah Arafat and Bi, Yu},
  journal={arXiv preprint arXiv:2409.01952},
  year={2024}
}

@article{miah2026badsnn,
  title={BadSNN: Backdoor Attacks on Spiking Neural Networks via Adversarial Spiking Neuron},
  author={Miah, Abdullah Arafat and Vu, Kevin and Bi, Yu},
  journal={arXiv preprint arXiv:2602.07200},
  year={2026}
}

@article{yang2024sampdetox,
  title={Sampdetox: Black-box backdoor defense via perturbation-based sample detoxification},
  author={Yang, Yanxin and Jia, Chentao and Yan, DengKe and Hu, Ming and Li, Tianlin and Xie, Xiaofei and Wei, Xian and Chen, Mingsong},
  journal={Advances in Neural Information Processing Systems},
  volume={37},
  pages={121236--121264},
  year={2024}
}

@inproceedings{fu2024spikewhisper,
  title={Spikewhisper: Temporal spike backdoor attacks on federated neuromorphic learning over low-power devices},
  author={Fu, Hanqing and Li, Gaolei and Wu, Jun and Li, Jianhua and Zhou, Kai and Liu, Yuchen},
  booktitle={International Conference on Neural Information Processing},
  pages={243--258},
  year={2024},
  organization={Springer}
}

@article{fang2021deep,
  title={Deep residual learning in spiking neural networks},
  author={Fang, Wei and Yu, Zhaofei and Chen, Yanqi and Huang, Tiejun and Masquelier, Timoth{\'e}e and Tian, Yonghong},
  journal={Advances in neural information processing systems},
  volume={34},
  pages={21056--21069},
  year={2021}
}

@inproceedings{abad2025time,
  title={Time-distributed backdoor attacks on federated spiking learning},
  author={Abad, Gorka and Picek, Stjepan and Urbieta, Aitor},
  booktitle={European Symposium on Research in Computer Security},
  pages={1--20},
  year={2025},
  organization={Springer}
}

@article{shi2023black,
  title={Black-box backdoor defense via zero-shot image purification},
  author={Shi, Yucheng and Du, Mengnan and Wu, Xuansheng and Guan, Zihan and Sun, Jin and Liu, Ninghao},
  journal={Advances in Neural Information Processing Systems},
  volume={36},
  pages={57336--57366},
  year={2023}
}

@inproceedings{xi2021graph,
  title={Graph backdoor},
  author={Xi, Zhaohan and Pang, Ren and Ji, Shouling and Wang, Ting},
  booktitle={30th USENIX security symposium (USENIX Security 21)},
  pages={1523--1540},
  year={2021}
}

@inproceedings{li2021hidden,
  title={Hidden backdoors in human-centric language models},
  author={Li, Shaofeng and Liu, Hui and Dong, Tian and Zhao, Benjamin Zi Hao and Xue, Minhui and Zhu, Haojin and Lu, Jialiang},
  booktitle={Proceedings of the 2021 ACM SIGSAC conference on computer and communications security},
  pages={3123--3140},
  year={2021}
}

@article{gu2017badnets,
  title   = {{BadNets}: Identifying Vulnerabilities in the Machine Learning Model Supply Chain},
  author  = {Gu, Tianyu and Dolan-Gavitt, Brendan and Garg, Siddharth},
  journal = {arXiv preprint arXiv:1708.06733},
  year    = {2017},
  url     = {https://arxiv.org/abs/1708.06733}
}

@article{orchard2015converting,
  title     = {Converting Static Image Datasets to Spiking Neuromorphic Datasets Using Saccades},
  author    = {Orchard, Garrick and Jayawant, Ajinkya and Cohen, Gregory K. and Thakor, Nitish V.},
  journal   = {Frontiers in Neuroscience},
  volume    = {9},
  pages     = {437},
  year      = {2015},
  publisher = {Frontiers}
}

@article{abad2023sneaky,
  title={Sneaky spikes: Uncovering stealthy backdoor attacks in spiking neural networks with neuromorphic data},
  author={Abad, Gorka and Ersoy, Oguzhan and Picek, Stjepan and Urbieta, Aitor},
  journal={arXiv preprint arXiv:2302.06279},
  year={2023}
}

@article{eshraghian2023training,
  title={Training spiking neural networks using lessons from deep learning},
  author={Eshraghian, Jason K and Ward, Max and Neftci, Emre O and Wang, Xinxin and Lenz, Gregor and Dwivedi, Girish and Bennamoun, Mohammed and Jeong, Doo Seok and Lu, Wei D},
  journal={Proceedings of the IEEE},
  volume={111},
  number={9},
  pages={1016--1054},
  year={2023},
  publisher={IEEE}
}

@article{serrano2013dvs,
  title={A 128$\times$128 1.5\% contrast sensitivity 0.9\% FPN 3 $\mu$s latency 4 mW asynchronous frame-free dynamic vision sensor using transimpedance preamplifiers},
  author={Serrano-Gotarredona, Teresa and Linares-Barranco, Bernab{\'e}},
  journal={IEEE Journal of Solid-State Circuits},
  volume={48},
  number={3},
  pages={827--838},
  year={2013},
  publisher={IEEE}
}

@inproceedings{zeng2023narcissus,
  title={Narcissus: A practical clean-label backdoor attack with limited information},
  author={Zeng, Yi and Pan, Minzhou and Just, Hoang Anh and Lyu, Lingjuan and Qiu, Meikang and Jia, Ruoxi},
  booktitle={Proceedings of the 2023 ACM SIGSAC Conference on Computer and Communications Security},
  pages={771--785},
  year={2023}
}

@inproceedings{zhao2020clean,
  title={Clean-label backdoor attacks on video recognition models},
  author={Zhao, Shihao and Ma, Xingjun and Zheng, Xiang and Bailey, James and Chen, Jingjing and Jiang, Yu-Gang},
  booktitle={Proceedings of the IEEE/CVF conference on computer vision and pattern recognition},
  pages={14443--14452},
  year={2020}
}

@article{yang2021careful,
  title={Be careful about poisoned word embeddings: Exploring the vulnerability of the embedding layers in nlp models},
  author={Yang, Wenkai and Li, Lei and Zhang, Zhiyuan and Ren, Xuancheng and Sun, Xu and He, Bin},
  journal={arXiv preprint arXiv:2103.15543},
  year={2021}
}

@article{chen2020event,
  title={Event-based neuromorphic vision for autonomous driving: A paradigm shift for bio-inspired visual sensing and perception},
  author={Chen, Guang and Cao, Hu and Conradt, Jorg and Tang, Huajin and Rohrbein, Florian and Knoll, Alois},
  journal={IEEE Signal Processing Magazine},
  volume={37},
  number={4},
  pages={34--49},
  year={2020},
  publisher={IEEE}
}

@inproceedings{viale2021carsnn,
  title={CarSNN: An efficient spiking neural network for event-based autonomous cars on the Loihi neuromorphic research processor},
  author={Viale, Alberto and Marchisio, Alberto and Martina, Maurizio and Masera, Guido and Shafique, Muhammad},
  booktitle={2021 International Joint Conference on Neural Networks (IJCNN)},
  pages={1--10},
  year={2021},
  organization={IEEE}
}

@article{kasabov2014evolving,
  title={Evolving spiking neural networks for personalised modelling, classification and prediction of spatio-temporal patterns with a case study on stroke},
  author={Kasabov, Nikola and Feigin, Valery and Hou, Zeng-Guang and Chen, Yixiong and Liang, Linda and Krishnamurthi, Rita and Othman, Muhaini and Parmar, Priya},
  journal={Neurocomputing},
  volume={134},
  pages={269--279},
  year={2014},
  publisher={Elsevier}
}

@inproceedings{doan2021lira,
  title={LIRA: Learnable, imperceptible and robust backdoor attacks},
  author={Doan, Khoa and Lao, Yingjie and Zhao, Weijie and Li, Ping},
  booktitle={Proceedings of the IEEE/CVF International Conference on Computer Vision},
  pages={11966--11976},
  year={2021}
}

@inproceedings{abad2024sneaky,
  title={Sneaky Spikes: Uncovering stealthy backdoor attacks in spiking neural networks with neuromorphic data},
  author={Abad, Gorka and Ersoy, Oguzhan and Picek, Stjepan and Urbieta, Aitor},
  booktitle={Proceedings of the Network and Distributed System Security Symposium (NDSS)},
  year={2024}
}

@article{li2025tmpbd,
  title={Unsupervised backdoor detection and mitigation for spiking neural networks},
  author={Li, Jiachen and Wu, Bang and Xia, Xiaoyu and Liu, Xiaoning and Yi, Xun and Zhang, Xiuzhen},
  journal={arXiv preprint arXiv:2510.06629},
  year={2025}
}

@article{kingma2014adam,
  title={Adam: A method for stochastic optimization},
  author={Kingma, Diederik P and Ba, Jimmy},
  journal={arXiv preprint arXiv:1412.6980},
  year={2014}
}

@inproceedings{liu2020reflection,
  title={Reflection backdoor: A natural backdoor attack on deep neural networks},
  author={Liu, Yunfei and Ma, Xingjun and Bailey, James and Lu, Feng},
  booktitle={European Conference on Computer Vision},
  pages={182--199},
  year={2020},
  organization={Springer}
}

@article{riano2024flashy,
  title={Flashy Backdoor: Real-world environment backdoor attack on SNNs with DVS cameras},
  author={Ria{\~n}o, Roberto and Abad, Gorka and Picek, Stjepan and Urbieta, Aitor},
  journal={arXiv preprint arXiv:2411.03022},
  year={2024}
}

@inproceedings{wang2019nc,
  title={Neural cleanse: Identifying and mitigating backdoor attacks in neural networks},
  author={Wang, Bolun and Yao, Yuanshun and Shan, Shawn and Li, Huiying and Viswanath, Bimal and Zheng, Haitao and Zhao, Ben Y},
  booktitle={2019 IEEE Symposium on Security and Privacy (SP)},
  pages={707--723},
  year={2019},
  organization={IEEE}
}

@inproceedings{liu2018finepruning,
  title={Fine-pruning: Defending against backdooring attacks on deep neural networks},
  author={Liu, Kang and Dolan-Gavitt, Brendan and Garg, Siddharth},
  booktitle={International Symposium on Research in Attacks, Intrusions, and Defenses},
  pages={273--294},
  year={2018},
  organization={Springer}
}

@inproceedings{li2021nad,
  title={Neural attention distillation: Erasing backdoor triggers from deep neural networks},
  author={Li, Yige and Lyu, Xixiang and Koren, Nodens and Lyu, Lingjuan and Li, Bo and Ma, Xingjun},
  booktitle={International Conference on Learning Representations},
  year={2021}
}

\end{document}